\documentclass[journal]{IEEEtran}
\usepackage{amsmath,amsfonts}
\usepackage{algorithmic}
\usepackage{algorithm}
\usepackage{array}
\usepackage[caption=false,font=footnotesize,labelfont=rm,textfont=rm]{subfig}
\usepackage{textcomp}
\usepackage{stfloats}
\usepackage{url} 
\usepackage{verbatim}
\usepackage{graphicx}
\usepackage{cite}
\usepackage{url}
\usepackage{amsmath}
\usepackage{amsthm}
\usepackage{amssymb}
\usepackage{upgreek}
\usepackage{algorithm}
\usepackage{algorithmic}
\usepackage{bm}
\usepackage[utf8]{inputenc}
\usepackage{diagbox} 
\usepackage{xcolor}
\usepackage{multirow}
\usepackage{booktabs}
\usepackage{tabularx}
\usepackage{cleveref}

\newtheorem{lemma}{Lemma}
\newtheorem{proposition}{Proposition}
\newtheorem{definition}{Definition}
\newtheorem{collary}{Collary}
\newtheorem{remark}{Remark}
\newtheorem{assumption}{Assumption}
\def\BibTeX{{\rm B\kern-.05em{\sc i\kern-.025em b}\kern-.08em
    T\kern-.1667em\lower.7ex\hbox{E}\kern-.125emX}}
\usepackage{balance}

\begin{document}

\title{SMDP-Based Dynamic Batching for Improving Responsiveness and Energy Efficiency of Batch Services}

\author{Yaodan~Xu,~\IEEEmembership{Student Member,~IEEE,}
Sheng~Zhou,~\IEEEmembership{Senior Member,~IEEE,} and~Zhisheng~Niu,~\IEEEmembership{Fellow,~IEEE}
\thanks{This work was supported in part by the National Natural Science Foundation of China under Grants 62341108 and in part by the Fundamental Research Funds for the Central Universities under Grant 2242022k60006.}
\thanks{Y. Xu, S. Zhou, and Z. Niu are with Department of Electronic Engineering, the Beijing National Research Center for Information Science and Technology (BNRist), Tsinghua University, Beijing 100084, China (e-mail: xyd21@mails.tsinghua.edu.cn; sheng.zhou@tsinghua.edu.cn; niuzhs@tsinghua.edu.cn).}
\thanks{Part of this work has been published in IEEE ICC 2023\cite{smdpxu23}.}}



\maketitle

\begin{abstract}
For servers incorporating parallel computing resources, batching is a pivotal technique for providing efficient and economical services at scale.
Parallel computing resources exhibit heightened computational and energy efficiency when operating with larger batch sizes.
However, in the realm of online services, the adoption of a larger batch size may lead to longer response times. 
This paper aims to provide a dynamic batching scheme that delicately balances latency and efficiency. 
The system is modeled as a batch service queue with size-dependent service times. 
Then, the design of dynamic batching is formulated as a semi-Markov decision process (SMDP) problem, with the objective of minimizing the weighted sum of average response time and average power consumption.
A method is proposed to derive an approximate optimal SMDP solution, representing the chosen dynamic batching policy. 
By introducing an abstract cost to reflect the impact of ``tail" states, the space complexity and the time complexity of the procedure can decrease by 63.5\% and 98\%, respectively.
Numerical results showcase the superiority of SMDP-based batching policies across various parameter setups. 
Additionally, the proposed scheme exhibits noteworthy flexibility in balancing power consumption and latency.

\end{abstract}


\section{Introduction}\label{sec:introduction}
\IEEEPARstart{T}{o} meet the escalating demands for  powerful computing capabilities, processors have undergone significant advancements in recent decades.
The processors of today, including multi-core processors, graphics processing units (GPUs) and tensor processing units (TPUs), are equipped to better support parallel computing.
This enhancement is crucial for efficiently managing large-scale data and executing complex tasks.
For instance, GPUs have played a prominent role in accelerating the training and inference of neural networks due to their advantage in parallel computing\cite{oh2004gpu}.
These computing resources are now widely deployed across various levels—locally, on edge servers, and on cloud servers, providing computing services that facilitate ubiquitous access to intelligence at any time and from anywhere.

\begin{figure}
    \centering    
    \includegraphics[width=0.72\linewidth]{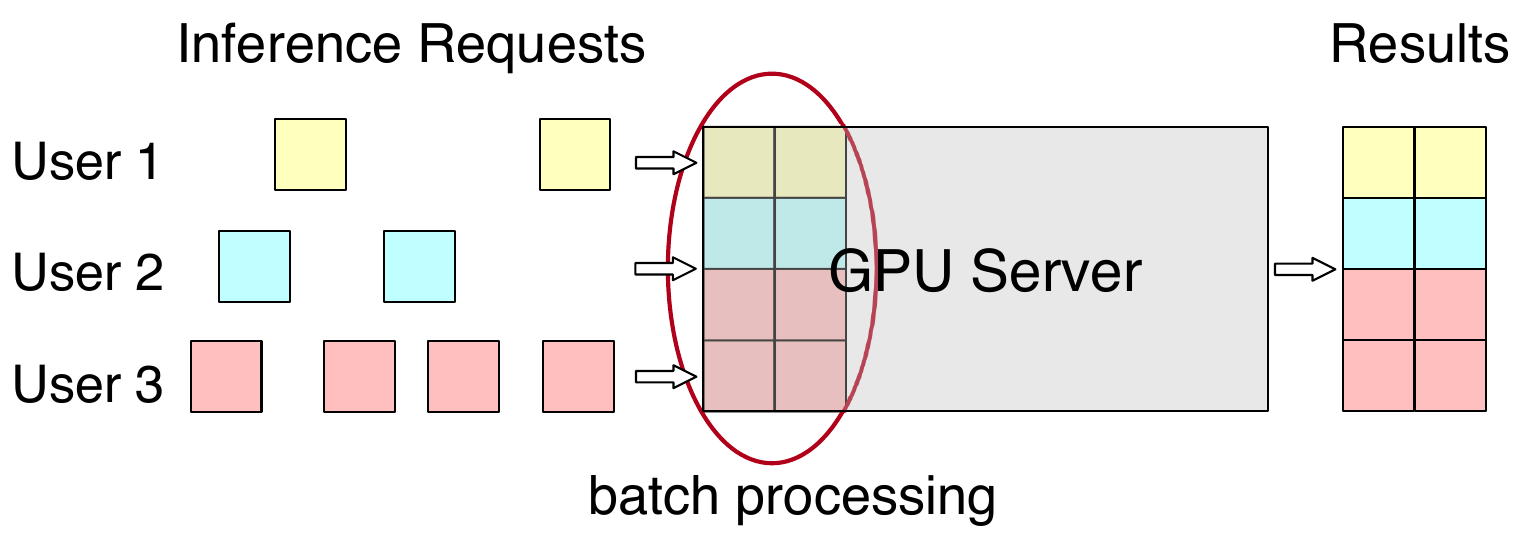}
    \vspace{-3mm}
    \caption{Batching of the same type of inference requests from potentially different users on a GPU-based ML-as-a-Service (MLaaS) platform.}
    \label{fig:batching}
\end{figure}

For servers equipped with processors capable of parallel operations, an important factor that affects both performance and cost of computing services is batch processing, or \emph{batching}\cite{zhang2019mark,crankshaw2017clipper}.
Specifically, batching is usually employed for homogeneous tasks that share common operations, allowing the grouping of data into a unified batch for simultaneous processing on the server.
The number of standard units of data or tasks gathered in a batch is referred to as the \emph{batch size}.
It is noteworthy that batching helps to better utilize computing and energy resources due to parallelism.
This batching effect has been examined across different hardware platforms for various tasks, including basic matrix computations\cite{shin2018mcdram} and diverse machine learning (ML) inference models\cite{wang2019benchmarking,NVIDIA,yao2022eais,wang2022energy,bhardwaj2023packrat,nabavinejad2022coordinated}.

However, opting for a larger batch size is not always the preferred approach, especially for online service provisioning where requests arrive in a random pattern, and expect responsive feedback.
This gives rise to two main challenges in determining the batch size:
(1) Larger batch sizes enhance energy efficiency and throughput by improving resource utilization and reducing per-sample I/O overhead. However, this benefit may come at the cost of decreased responsiveness, as batching can increase request response time due to potential waiting time needed to form a batch and extended processing time for handling multiple requests simultaneously\cite{ali2020batch,crankshaw2017clipper,zhang2023cofb}. This creates a \emph{tradeoff} between efficiency and responsiveness\cite{ali2020batch}.
(2) The use of statically configured batching proves inadequate in realistic scenarios\cite{choi2021lazy}, exhibiting poor responsiveness under low load conditions and limited throughput under high load\cite{cui20202}.
To address these issues, a dynamic batching scheme is essential, allowing for judicious batch size adjustments to well balance the performance and cost.

In this paper, we study the dynamic batching scheme on batch processing-capable servers, aiming to strike a delicate balance between responsiveness and energy efficiency. 
This issue has gained increasing significance with the emergence of ML-as-a-Service (MLaaS) platforms like Google Cloud Prediction\cite{GoogleCloud}, where trained ML models are published on the platforms to provide inference (prediction) services for massive end users.
As illustrated in Fig.~\ref{fig:batching}, batch processing the inference requests becomes a natural strategy, for efficiency and economical concerns.

We commence our exploration with intra-processor parallelism, considering a scenario with a single server equipped with a single parallel computing processor. 
We leverage the theoretical framework of sequential decision-making to address the sequential batch size decisions.
In our context, where we assume Poisson request arrivals and an arbitrary service time distribution, the problem is formulated as a semi-Markov decision process (SMDP).
The objective is a weighted sum of the long-term average request response time and average power consumption. 
Notably, to the best of our knowledge, the optimal control problem for batch service queues with size-dependent service times remains unexplored in the literature. 
Moreover, the inherent complexities of the formulated SMDP problem---characterized by an infinite state space, an average (non-discounted) objective, and unbounded costs---pose challenges for efficient resolution using traditional methods.
To address these challenges, we propose a procedure to solve the SMDP problem and derive an approximate optimal policy, which manifests as the selected dynamic batching scheme.
Our main contributions are summarized as follows:
\begin{enumerate}
    \item 
    To the best of our knowledge, our work is the first to rigorously formulate and optimally solve the dynamic batching problem for online computing services.
    The batching decision is formulated as an infinite-state SMDP, with the objective of minimizing the weighted sum of average response time and average power consumption.
    
    \item 
    A new method is proposed, composed of finite state approximation, model ``discretization" and relative value iteration, to obtain an approximate optimal policy.
    The demanding problem of state space explosion is tackled by a novel abstract cost, which reflects the impact of costs in ``tail" states. 

    \item 
    We also conclude the theoretical results regarding the optimal policy structure in special cases.
    The SMDP solutions obtained through the proposed general method are visualized under different parameter settings. 
    On one hand, the computed policies align with the theoretical results in special cases, affirming the effectiveness of the proposed method and the correctness of the theoretical results. 
    On the other hand, certain instances reveal that the theoretical results might not extend to more general scenarios, underscoring the necessity of the general solving approach. 

    \item
    Extensive numerical results demonstrate that the SMDP-based policies achieve the lowest average cost compared to benchmarks. 
    The latency-energy tradeoff curves show that when having the same average response time, the SMDP-based policies never consume more energy than any other benchmark policy, and vice versa.
    Moreover, the proposed scheme can adapt to different traffic intensities and flexibly balance the response time and power consumption.
\end{enumerate}

\begin{figure}[h]
    \centering
    \subfloat[Batch inference latency and batch service rate.]{%
        \includegraphics[width=0.48\textwidth]{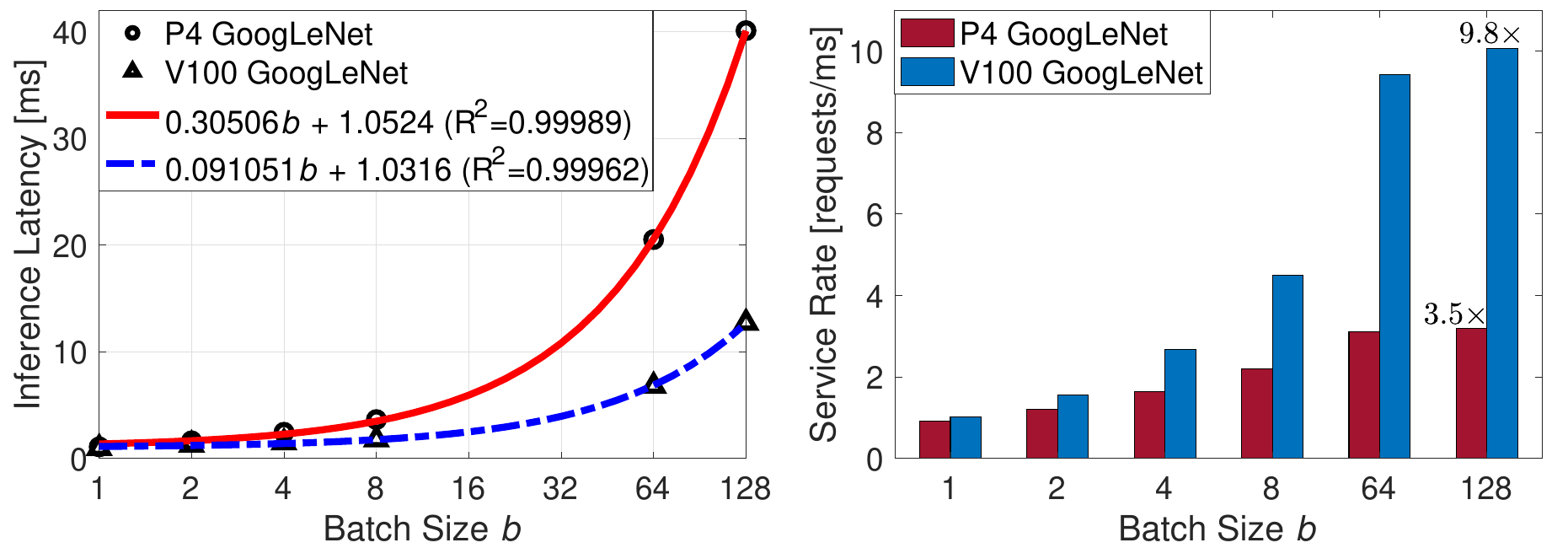}%
        \label{fig:latency}%
    }
    \vspace{0mm} 
    \subfloat[Batch energy consumption and energy efficiency.]{%
        \includegraphics[width=0.48\textwidth]{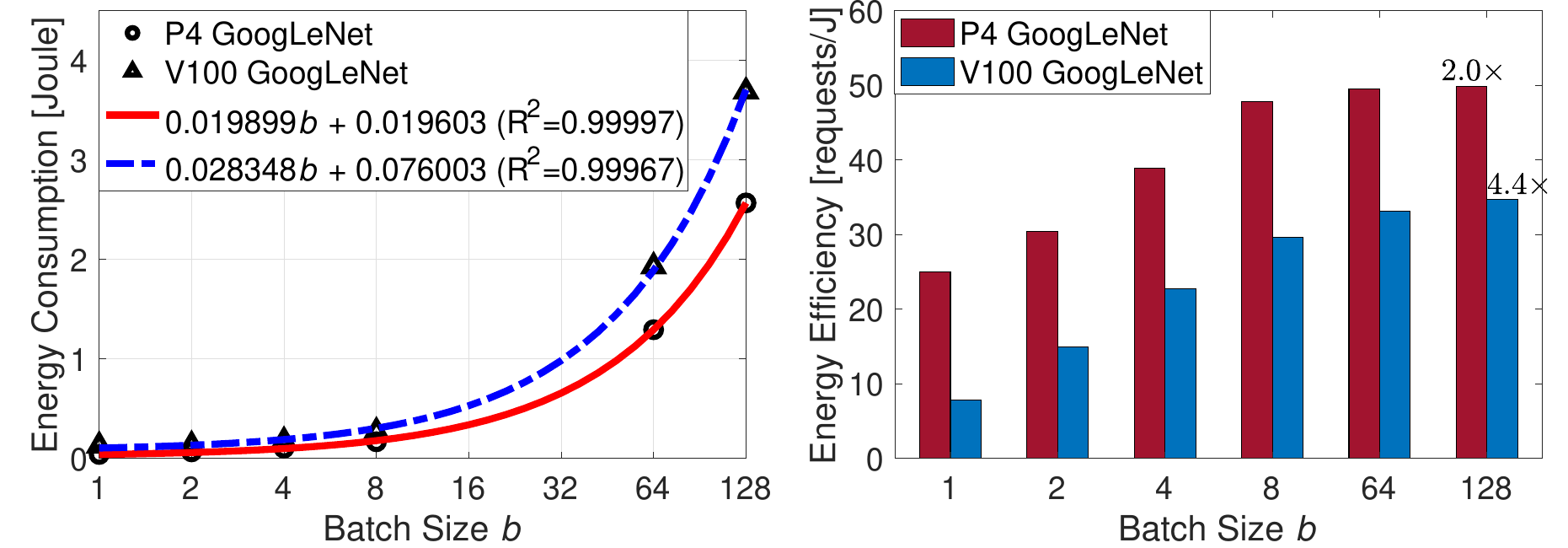}%
        \label{fig:energy}%
    }
    \caption{Inference latency and energy consumption for batch processing GoogLeNet\cite{szegedy2015going} on TESLA P4 and TESLA V100.
    The data, measured by NVIDIA, is based on an image classification task using images from the ImageNet12 dataset, which consists of 1000 classes with an image size of $224 \times 224$\cite{NVIDIA}. 
    The batch size is plotted in $\log_2$ coordinate.}
    \label{fig:twosub}
\end{figure}

The rest of this paper is organized as follows.
We begin by presenting related works in Section~\ref{sec:related_works}.
The system model is introduced in Section~\ref{sec:system2}, followed by the SMDP formulation in Section~\ref{sec:formulation}.
A procedure for solving the SMDP problem is proposed in Section~\ref{section:procedure}.
Theoretical analyses regarding the optimal policy in special cases of the problem are detailed in Section~\ref{sec:special cases}.
Numerical results are showcased in Section~\ref{sec:results}.
Finally, Section~\ref{sec:conclusion} provides the concluding remarks for the paper.

\section{Related Works}\label{sec:related_works}

\textbf{Dynamic Batching.}
Some works have explored dynamic batching for systems such as data centers\cite{wang2012virtual} and Spark Streaming\cite{cheng2018adaptive}.
Recently, there has been a notable increase in attention towards dynamic batching, particularly in the context of ML applications.
While works like DBS\cite{ye2020dbs} and Zeus\cite{you2023zeus} explore batching for ML training, they fall outside the scope of our focus. 
Our exclusive attention is on online services. 
For example, in applications such as smart healthcare monitoring, data generated by devices can be efficiently processed through cloud, edge, or hybrid computing solutions to ensure timely responses\cite{yadav2020energy,ling2024qos,yadav2021smart}.
Plenty of research has been conducted on dynamic batching for ML inference serving on GPU-based platforms\cite{crankshaw2017clipper,  zhang2019mark, cui20202, choi2021lazy, ali2020batch, yao2022eais,serf,li2023alpaserve,qin2019swift}.
There are also studies addressing the batching issue on a multi-core central processing unit (CPU)\cite{bhardwaj2023packrat}.
Nevertheless, progress in theoretical analysis remains very limited.
For example, SERF\cite{serf} models inference serving as an M/D/c queue. However, it does not explicitly account for batching in the model.
Another work, BATCH\cite{ali2020batch}, characterizes request arrivals as a Poisson process or a Markov-modulated Poisson
process with two phases (MMPP(2))\cite{fischer1993markovmodulated}, allowing for the estimation of the number of requests that arrive before a timeout.
However, this analysis overlooks the cases where requests arrive during processing times.
In contrast, the author of \cite{inoue2021queueing} presents a closed-form queueing analysis of dynamic batching under the greedy batching policy. 
Nonetheless, this policy is suboptimal as it does not leverage the potential benefits of larger batches.
Additionally, the analysis in \cite{inoue2021queueing} assumes an infinite batching capacity, which is generally impractical.

\textbf{Parallel Batch Processing.}
Parallel batch processing, or parallel batching, is a classical issue in operational research that finds applications in numerous fields such as manufacturing, transportation, and healthcare\cite{bonvin2006control,sasikala2016bulk,fowler2021survey}.
However, the batch processing problem studied in this paper exhibits two distinctive deviations from classical scenarios in the existing literature.
Firstly, unlike the ideal parallelism with batch-size independent service times\cite{deb1973optimal,zeng2017optimal}, the batch processing time can increase with the batch size\cite{shi2022multi, hanhirova2018latency}.
Secondly, the energy efficiency can directly benefit from an increased batch size, rather than remaining unchanged. 
For instance, as illustrated in Fig.~\ref{fig:twosub}\subref{fig:latency} and Fig.~\ref{fig:twosub}\subref{fig:energy}, which are based on the statistics from NIVIDIA\cite{NVIDIA}, the processing time and energy consumption for batch ML inference services appear to be affine functions of the batch size.
Consequently, the average processing time and energy consumption per batch decrease as the batch size increases, leading to improvements in both computational and energy efficiency, as shown in Fig.~\ref{fig:twosub}.
For GoogLeNet inference on a TESLA V100, using a batch size of 128 can achieve a speedup of 9.8 times compared to inference without batching (batch size of 1).
Additionally, the energy efficiency is improved by 4.4 times.
Queueing analyses for batch servers with size-dependent service times have been conducted in the literature, but they only focus on certain structured policies that are \emph{suboptimal}\cite{neuts1967general,maity2015analysis, pradhan2020distribution,gupta2021analysis}.
Following the line of optimal control, existing research\cite{deb1973optimal,papadaki2002exploiting,fowler2021survey} primarily addresses problems where batch processing times are \emph{independent} of the batch size.
In \cite{inoue2022load}, the author models the load-balancing problem in multiple batch service queues with \emph{size-dependent} processing times as a Markov decision process and identifies it as \emph{an open problem}.
In fact, optimal batching in one such batch service queue, as the simplest sub-problem of \cite{inoue2022load}, still remains unsolved.

\textbf{SMDP.}
Continuous-time Markov decision processes (CTMDPs) and SMDPs are the common formulations for sequential decision-making in continuous-time systems\cite{puterman1994markov}.
Given that the processing times of computation tasks often exhibit limited randomness\cite{ali2020batch}, deviating from the characteristics of an exponential distribution, SMDPs appear as more fitting choices for the studied problem.
In this work, we need to address an infinite-horizon and infinite-state SMDP problem with unbounded costs, where the objective is expressed in a long-term average form. 
Analytical results regarding optimal policies for this SMDP are available only in specific cases\cite{deb1973optimal}. 
An alternative approach is to utilize iteration-based numerical methods, which require finite state approximation due to the intractable infinite state space.
In prior research, the authors in \cite{thomas1985finite} demonstrated the convergence of several finite state approximation algorithms for average cost SMDP models, but only with bounded costs. 
In \cite{white1982finite}, proofs were provided for the convergence of finite state approximation algorithms for models with unbounded costs, but with a discounted objective. 
Nevertheless, none of the existing finite state approximation algorithms has been proven effective for the considered SMDP problem.

In summary, the lack of theoretical guarantees in existing dynamic batching schemes necessitates a rigorous formulation. 
The extracted system model distinguishes from classical batch service queues in its size-dependent service times, a feature that remains unexplored in the literature. 
The design of dynamic batching can be formulated into an SMDP problem with an infinite state space.
Regrettably, none of the existing finite state approximation algorithms has been proven effective in addressing the studied SMDP, highlighting the need for the development of novel methods.

In our prior work\cite{smdpxu23}, we proposed an SMDP-based dynamic batching scheme for ML inference serving on GPU-based platforms. 
This paper extends the research to encompass general online computing service scenarios. 
The batch service time and energy consumption take more general forms with respect to the batch size, unlike the deterministic and linear functions considered in \cite{smdpxu23}. 
Moreover, additional insights are gained through both theoretical induction and extensive numerical results.
\section{System Model}\label{sec:system2}
We consider a single server with a single parallel computing processor in the continuous-time setting.
Batch processing is implemented on the same type of computing requests, and it cannot be interrupted once the processing is started.
The system is modeled as a single service queue, where computing requests (tasks) are assumed to arrive according to a Poisson process with an arrival rate of $\lambda$. 
The requests awaiting processing are stored in a buffer, which is assumed to have an infinite capacity.
This assumption of an infinite buffer is based on the fact that the storage capacity of a computing server, which functions as the buffer in the queuing model, is significantly larger than the memory size.
The total number of requests in the buffer as well as being processed at time $t \geq 0$ is denoted by $s(t) \in \{0,1,2,\dots\}$.

Let $b \in \mathcal{B} \triangleq \{x \, | \, x \in \mathbb{N}_{+}, B_{\min} \leq x \leq B_{\max}\}$ denote the batch size, where $ B_{\min}, B_{\max} \in \mathbb{N}_{+}$ and $ B_{\min} \leq B_{\max}$.
$B_{\max}$ (or $B_{\min}$) is the maximum (or minimum) batch size allowed by the system.
The cumulative distribution function (CDF) of the processing (service) time for a batch of size $b$ is $G_{b}(x) \; (x \geq 0)$ with mean $1/\mu^{[b]}$.
Assume that for every $b \in \mathcal{B}$, $G_b(x)$ has a finite second moment.
Moreover, the mean processing time should be finite and larger than zero, i.e., $G_{b}(0)<1$ and $1/\mu^{[b]} < \infty$.
Let $l \; : \; \mathcal{B} \rightarrow \mathbb{R}$ represent the function of the mean batch processing time with respect to the batch size, where for $\forall b \in \mathcal{B}$, $0<l(b)<\infty$ and ${l(b)} = \frac{1}{\mu^{[b]}}$.
The time required for processing a larger batch should be no less than that for a smaller batch. 
Therefore, we assume that $l(b)$ is monotonically non-decreasing in terms of $b$.
Let $\tilde{g}_{b}(s)$ denote the Laplace transform of the service time distribution $G_{b}(x)$, defined by
\begin{equation}
    \tilde{g}_{b}(s)=\int_{0}^{\infty}e^{-sx}\mathrm{d}G_{b}(x),
\end{equation}
with $-\tilde{g}^{\prime}_{b}(0)=1/\mu^{[b]}$.

When operating with a batch size of $b$,  define the batch processing computational efficiency, or batch service rate, as the average number of requests processed per unit of time, denoted as $\theta(b) \triangleq \frac{b}{l(b)}$.
Since parallelism can enhance computational efficiency, it is assumed that $\theta(b)$ is monotonically non-decreasing with the batch size $b$.
Noting that $l(b)$ is also a non-decreasing function with respect to $b$, it follows that the mean batch processing time $l(b)$ should exhibit \emph{linear} or \emph{sublinear growth} as $b$ increases, if $l(b)$ is dependent on $b$.

\begin{remark}
    \textbf{(i)} When $l(b)$ exhibits \textbf{sublinear} growth as $b$ increases, the computational efficiency $\theta(b)$  \textbf{monotonically increases} with $b$. 
    \textbf{(ii)} When $l(b)$ exhibits \textbf{linear} growth as $b$ increases, there are two cases: 
    If $l(b)$ is \textbf{proportional} to $b$, i.e., $l(b)= \alpha b$ with $\alpha > 0$, the computational efficiency $\theta(b)$ is \textbf{constant} and independent of $b$; 
    If $l(b)$ is \textbf{affine} to $b$, i.e., $l(b)= \alpha b + l_0$ with $\alpha > 0$ and $l_0 > 0$, the computational efficiency $\theta(b)$ \textbf{monotonically increases} with $b$. 
\end{remark}

Thus, the maximum service rate is $\theta(B_{\max})=\frac{B_{\max}}{l(B_{\max})}={B_{\max}}\mu^{[B_{\max}]}$.
Let $\rho=\lambda/(B_{\max}\mu^{[B_{\max}]})$ denote the ratio of the arrival rate over the maximum service rate.
Assume that $\lambda l(B_{\max}) < B_{\max}$, or equivalently $\lambda < B_{\max}\mu^{[B_{\max}]}$, then $\rho$ satisfies $\rho \in (0,1)$, which is a necessary condition for
the system stability.

The energy consumption of processing a batch of $b$ requests is denoted by ${\zeta}(b)$, $\zeta \; : \; \mathcal{B} \rightarrow \mathbb{R}$.
Let $\eta(b)=\frac{b}{{\zeta}(b)}$ denote the batch processing energy efficiency, which is defined as the average number of requests served with one unit of energy consumption.
Given the potential for parallelism to improve energy efficiency, we assume that  $\eta(b)$ is monotonically non-decreasing with the batch size $b$.
Similarly, $\zeta(b)$, the function of the batch processing energy consumption, should
exhibit \emph{linear} or \emph{sublinear growth} as $b$ increases, if $\zeta(b)$ is not
independent of $b$.

Given the server configurations and the specific computing task, the parameters $B_{\max}$ and $B_{\min}$ can be profiled and determined.
Additionally, the exact forms of $G_b(\cdot), l(b)$, and ${\zeta}(b)$ are established by fitting the latency and energy consumption statistics obtained from
prior profiling\cite{ali2020batch, inoue2021queueing}.




We consider two main factors in the objective: One is the request response time (or latency), which includes both waiting and processing time, as the performance metric. 
The other is the power consumption of the server, as the running cost metric.
Our objective is to minimize the weighted sum of average request response time, denoted by $\overline{W}$, and average power consumption, denoted by $\overline{P}$:
\begin{equation}
    \min \; w_1\overline{W}
    + w_2\overline{P},
\label{eq:multi-obj}
\end{equation}
where $w_1 > 0$ and $w_2 \geq 0$ are the weights.

The serving process consists of sequential service rounds.
Define $t_i\; (t_i \geq 0, \;i=1,2,\ldots)$ as the start time of the $i$th service round and $b(t_i) \in \mathcal{B}$ as the batch size in the $i$th service round.
Let $N(t) \in \mathbb{N}$ denote the total number of service rounds until time $t \geq 0$.
The objective can then be expressed as
\begin{equation}
        \min \; \limsup\limits_{T\rightarrow\infty} \frac{1}{T}
        \Bigg\{
        w_1 \frac{1}{\lambda} \int_{0}^{T} s(t)  \mathrm{d}t
        +
        w_2\sum_{i=1}^{N(T)} {\zeta}{(b(t_i))}
        \Bigg\}
        ,
    \label{obj1}
\end{equation}
Note that here the average request response time $\overline{W}$ is equivalently transformed to the average queue length $\overline{L}$ through Little's Law $\overline{L}=\lambda\overline{W}$\cite{little1961proof}, where $\overline{L}=\limsup\limits_{T\rightarrow\infty} \frac{1}{T}\int_{0}^{T} s(t)\mathrm{d}t$.
    
\begin{remark}
The energy cost considered in this model can also be replaced with other types of costs, such as the monetary cost\cite{ali2020batch}.
\end{remark}

\section{SMDP Formulation}\label{sec:formulation}
The decision-making process for batching in the continuous-time system, as introduced in Section~\ref{sec:system2}, naturally lends itself to the formulation as an  SMDP\cite{puterman1994markov}.
In SMDP, we are only concerned with the states at decision epochs, upon which the decisions are made.
Let $\mathcal{T}=[0,\infty)$ be the timeline of the SMDP model.
The decision epochs are set as the moments when either the server completes a batch of service, or a request arrives while the server is idle.
The $m$th ($m=0,1,2,\ldots$) decision epoch is denoted as
$t_m \in \mathcal{T}$. 
Let the state be the number of requests in the system.
The state at the $m$th decision epoch is denoted by $s_m$, taking values from the state space $\mathcal{S} \triangleq \{0,1,2,\ldots\}$.


At each epoch $m$, the server takes an action $a_m$ from the
action space $\mathcal{A} \triangleq \{0\} \cup \mathcal{B}$.
The action $a_m$ is the size of batch to be processed.
Note that $a_m=0$ means that no requests are served in the $m$th epoch.
Let $\mathcal{A}_s \subseteq \mathcal{A}$ be the set of feasible actions for a given state $s$.
The number of requests to be batched should be no more than the available requests, which means $\mathcal{A}_s \triangleq \{0,1,2,\ldots,s\} \cap \mathcal{A}$, and thus $\mathcal{A}_s = \{0,1,2,\ldots,s\} \cap \mathcal{B}$.

The state transition is associated with the current state and action.
Let $m(j|s,a)$ denote the probability that the semi-Markov decision process occupies state $j$ at the next decision epoch when action $a$ is chosen at the current state $s$.
Let $p_k^{[b]}$ denote the probability that $k$ requests arrive during the period of processing a batch of $b \in \mathcal{B}$ requests.
With the assumption that the arrival of requests follows a Poisson process, we have
\begin{equation}
    p_k^{[b]} 
    = 
    \int_0^{\infty}
    \frac{e^{-\lambda t}(\lambda t)^k}{k !}
    \mathrm{d} G_b(t)
    , \quad k=0,1,2, \ldots
    \label{pkb}
\end{equation}
A useful method to generate the probabilities $p_k^{[b]}$ is by using $\tilde{g}_{b}(s)$, which is the Laplace transform of the service time distribution.
Denote the probability generating function (PGF) that corresponds to $p_k^{[b]}$ as $A^{[b]}(z)=\sum_{k=0}^{\infty}p_k^{[b]}z^k$, which can be simplified to
\begin{equation}
\begin{aligned}
A^{[b]}(z) 
&=\sum_{k=0}^{\infty} p_k^{[b]} z^k
=\sum_{k=0}^{\infty} z^k \int_0^{\infty} \frac{{e}^{-\lambda t}(\lambda t)^k}{k !} \mathrm{d} G_b(t)
\\
&=\int_0^{\infty} {e}^{-\lambda t}\left(\sum_{k=0}^{\infty} \frac{(\lambda t z)^k}{k !}\right) \mathrm{d} G_b(t) \\
& =\int_0^{\infty} {e}^{-\lambda t(1-z)} \mathrm{d} G_b(t)=\tilde{g}_b(\lambda(1-z)).
\end{aligned}
\end{equation}
Then the required probabilities $p_k^{[b]}$ can be computed by 
\begin{equation}
p_k^{[b]}=\left.\frac{1}{k !} \frac{\mathrm{d}^k \tilde{g}_b(\lambda(1-z))}{\mathrm{d} z^k}\right|_{z=0}, \quad k=0,1, \ldots
\label{eq6}
\end{equation}

The transition probability for $\forall j,s \in \mathcal{S}, \forall a \in \mathcal{A}_s$ is expressed as
\begin{equation}
m(j|s,a)=
\begin{cases}
p_{j-s+a}^{[a]}& \text{ $ j \geq s-a,$ $a \in \mathcal{A}_s,$ $a \neq 0$ } \\
1& \text{ $ j = s+1,$ $a=0$ }
\\
0& \text{ otherwise }
\end{cases}.
\label{eqp}
\end{equation}

Let a random variable $\gamma_m$ denote the sojourn time between the
$m$th and the $(m+1)$th epoch.
The random variables $\gamma_m, m \in \mathbb{N}$, are conditionally independent given the state $s_m$ and action $a_m$.
Let $\Gamma_{s,a}(x)\; (x \geq 0)$ denote the CDF of $\gamma_m$ when action $a \in \mathcal{A}_s$ is chosen at the state $s$, given by
\begin{equation}
    \Gamma_{s,a}(x) =
    \begin{cases}
   G_a(x)& \quad  a \in \mathcal{A}_s, a \neq 0
    \\
    1- e^{-\lambda x}&
    \quad  a = 0
    \end{cases},
    \; \forall s \in \mathcal{S}.
\end{equation}

Define $y(s, a)=\mathbb{E}[\gamma|s,a]$ as the expected sojourn time until the next decision epoch, given by
\begin{equation}
    y(s,a) =
    \begin{cases}
   1/{\mu}^{[a]}& \quad  a \in \mathcal{A}_s, a \neq 0
    \\
    1/\lambda&
    \quad  a = 0
    \end{cases},
    \; \forall s \in \mathcal{S}.
    \label{eqy}
\end{equation}

Costs are incurred for serving the requests as well as holding them.
The cost of serving a batch of $n \in \mathbb{N}$ requests is denoted by $u(n)$, and the cost of holding $n$ requests in the system \emph{per unit time} is denoted by $v(n)$.
Let $c(s, a)$ denote the expected cost until the next decision epoch when action $a \in \mathcal{A}_s$ is taken in state $s$. 
We have $c(s,0)=u(0)+v(s)y(s,0)$, and for $a \neq 0$,
\begin{equation}
c(s,a)=
  u(a)+
\int_0^{\infty}\int_0^{x}
\sum_{k=0}^{\infty} v(s+k) \frac{e^{-\lambda t}(\lambda t)^k}{k !} \mathrm{d} t
\mathrm{d} \Gamma_{s,a}(x).
\end{equation}

The cost functions corresponding to the objective in (\ref{obj1}) are $u(n)=w_2 {\zeta}(n) \; (n>0)$, $u(0)=0$  and $v(n)=\frac{w_1}{\lambda} n$.
This leads to a detailed description of $c(s,a)$, which is
\begin{equation}
\begin{aligned}
    c(s,0) &= w_1\frac{s}{{\lambda}^2},
    \\
    c(s,a) &=
    w_2{\zeta}(a) +
\int_0^{\infty}
\int_0^{x} \frac{w_1}{\lambda}(s+\lambda t)
    \mathrm{d} t
    \mathrm{d} \Gamma_{s,a}(x)
    \\
    &= 
    w_2{\zeta}(a) +
    \int_0^{\infty}
    {w_1}(\frac{s}{\lambda}x+\frac{1}{2}{x}^2)
    \mathrm{d} \Gamma_{s,a}(x)
    \\
    &=
    w_2{\zeta}(a) +
    {w_1}(\frac{s}{\lambda}\mathbb{E}[\gamma|s,a]
    +\frac{1}{2}\mathbb{E}[\gamma^2|s,a])
    \\
    &= 
    w_2{\zeta}(a) +
    {w_1}(\frac{s}{\lambda}\mathbb{E}[G_a]
    +\frac{1}{2}\mathbb{E}[G_a^2])
    \\
    &= 
    w_2{\zeta}(a) +
    {w_1}(\frac{s}{\lambda {\mu}^{[a]}}
    +\frac{1}{2}\mathbb{E}[G_a^2]),
     \; a \in \mathcal{A}_s, a \neq 0,
\end{aligned}
\label{cost_step}
\end{equation}
where $G_a$ denotes a generic random variable that follows the CDF $G_a(x)$, and $\mathbb{E}[G_a^2]=\tilde{g}^{\prime\prime}_{b}(0)$.

We can also generalize $c(s,a)$ in the form 
\begin{equation}
    c(s,a) = u(a) + d(s,a)y(s,a),
\end{equation}
where $d(s,a)$, referred to as the cost rate, represents the holding cost averaged over the sojourn time:
\begin{equation}
    d(s,a) =
    \begin{cases}
    w_1\frac{s}{{\lambda}}& \quad a = 0
    \\
    w_1(\frac{s}{{\lambda}}+\frac{1}{2}\mathbb{E}[G_a^2]{\mu}^{[a]})&
    \quad  a \in \mathcal{A}_s, a \neq 0
    \end{cases},
    \forall s \in \mathcal{S}.
\end{equation}

The formulated SMDP can be fully described by the set of objects $\mathcal{P} \triangleq \{\mathcal{T},\mathcal{S},\mathcal{A}_s,m(\cdot|s,a), \Gamma_{s,a}(\cdot),c(s,a)\}$, and we will use the symbol $\mathcal{P}$ to represent this SMDP model in the subsequent text. 



\begin{remark}
    
    This formulation assumes Poisson arrivals but does not restrict the distribution of the processing time.

    To extend the SMDP method to more general arrival processes, ﬁctitious decision epochs and additional state variables are required to maintain the semi-Markov property.
    For example, with MMPP arrivals, phase shifts must be included as decision epochs. For non-memoryless renewal arrivals, such as deterministic processes, the arrival points should be incorporated as decision epochs, and the remaining time of the arrival interval must be included in the state. 
    Because of the decision epochs occurring during processing times, the remaining processing time must also be part of the state for both cases.
    Moreover, an extra state variable is needed to distinguish the exact event of the decision epoch.

    Handling continuous state variables and addressing the curse of dimensionality caused by the enlarged state space are significant challenges.
    Therefore, utilizing SMDP with both general arrival intervals and processing times is quite difficult.
\end{remark}

Let $\mathcal{B}(\mathcal{A})$ denote the set of probability distributions on Borel subsets of $\mathcal{A}$.
A Markovian decision rule  $\bm{d}^{(m)}: \mathcal{S} \rightarrow \mathcal{B}(\mathcal{A})$ specifies the probabilities of taking each action at epoch $m$. 
It is called \emph{Markovian} since it relies solely on the current state $s_m$ for its decision-making.
The decision rule is \emph{deterministic} if it selects an action with probability 1.
A policy $\bm{\pi}=\{\bm{d}^{(0)},\bm{d}^{(1)},\bm{d}^{(2)},\ldots \}$ is a sequence of decision rules.
Furthermore, $\bm{\pi}$ is called \emph{stationary} if $\bm{d}^{(m)}=\bm{d}, \forall m \in \mathbb{N}$.



Our goal is to find a policy that minimizes the long term average expected
cost $g^{\bm{\pi}}(s_0)$, given $s_0 \in \mathcal{S}$ at $t=0$, which is
\begin{equation}
   \min \limits_{\bm{\pi}}
   g^{\bm{\pi}}(s_0)=
   \limsup\limits_{M \rightarrow\infty} \frac
   {{\mathbb{E}}^{\bm{\pi}}_{s_0}
        \Big[
        \sum_{m=0}^{M} c(s_m, a_m)
        \Big]}
   {{\mathbb{E}}^{\bm{\pi}}_{s_0}
        \Big[
        \sum_{m=0}^{M} \gamma_m
        \Big]}.
\label{objective2}
\end{equation}
The objective (\ref{objective2}) and the SMDP model $\mathcal{P}$ together constitute the SMDP problem.
The objective focuses on the long-term average, and for every history-dependent policy, there exists an equivalent Markovian policy with the same objective value (referred to Theorem 8.1.2 in \cite{puterman1994markov}). 
This implies that only Markovian policies need to be considered.
Moreover, in this paper, we restrict our consideration to \emph{stationary deterministic} policies.
A stationary deterministic policy $\pi:\mathcal{S} \rightarrow \mathcal{A}$ is a function that maps the state space $\mathcal{S}$ to the action space $\mathcal{A}$.
This type of policy is concise and clear, helping to reduce the solution space. 
For instance, the static batching policy and the greedy batching policy are both stationary deterministic policies.

\begin{definition}
A static batching policy with a parameter $b \in \mathcal{B}$ is denoted as $\pi^{b}_{\mathrm{static}}  :  \mathcal{S} \rightarrow \mathcal{A}$, and is defined as follows: 
\begin{equation}
    \pi^{b}_{\mathrm{static}}(s) =
    \begin{cases}
   0& \quad  s < b
    \\
    b&
    \quad  s \geq b
    \end{cases}.
\end{equation}
Under such a policy $\pi^{b}_{\mathrm{static}}$, the served batches have a constant batch size of $b$.
\label{definition_static}
\end{definition}

The greedy batching policy is a representative dynamic batching policy, defined as follows:

\begin{definition}
Define a greedy batching policy $\pi_{\mathrm{greedy}}  :  \mathcal{S} \rightarrow \mathcal{A}$ as 
\begin{equation}
    \pi_{\mathrm{greedy}}(s) =
    \max\{\min\{s,B_{\max}\},B_{\min}\}.
\end{equation}
Under such a policy $\pi_{\mathrm{greedy}}$, the system greedily serves batches with the current maximum allowable sizes.
\label{definition_con}
\end{definition}

The considered model $\mathcal{P}$ is an infinite state SMDP with non-negative, unbounded costs and finite action sets.
The existence of an optimal stationary deterministic policy for such a model requires further discussions.

\begin{proposition}
An average expected optimal
stationary deterministic policy exists for the SMDP model $\mathcal{P}$.
\label{proposition1}
\end{proposition}

\begin{IEEEproof}
See Appendix~A.
\end{IEEEproof}

The equations corresponding to the optimal stationary
deterministic policies are provided as follows.

\begin{proposition}
Let $h: \mathcal{S} \rightarrow \mathbb{R}$ denote a value function, and let $g \in \mathbb{R}$ represent a scalar.
Given the SMDP model $\mathcal{P}$, the constant and functions $(g,h)$ such that
\begin{equation}
\begin{aligned}
 h(s)=\min _{a \in \mathcal{A}_s}\bigg\{c(s, a)-g y(s, a)+\sum_{j \in S} m(j | s, a) h(j)\bigg\},&   \\
 \forall s \in \mathcal{S},&
\end{aligned}
\label{eq7}
\end{equation}
are exactly the \textbf{optimal} average expected cost per unit time, and the corresponding relative value functions. 
The function $h$ is referred to as the relative value function since $h(s)$ is exactly the relative expected total cost when starting with state $s$.

Consequently, the optimal stationary deterministic policy ${\pi}^*$ for the SMDP problem is given by
$$
\begin{aligned}
\pi^{*}(s) \in \underset{a \in \mathcal{A}_s}{\arg \min }
\bigg\{c(s, a)-g y(s, a)+\sum_{j \in S} m(j | s, a) h(j)\bigg\},&
\\
\forall s \in \mathcal{S} .&
\end{aligned}
$$

Equations (\ref{eq7}) are referred to as the optimality equations for the SMDP problem with $\mathcal{P}$ and the average objective.
\label{proposition2}
\end{proposition}

\begin{IEEEproof}
See Appendix~B.
\end{IEEEproof}

Through Proposition \ref{proposition1}, the existence of an optimal stationary deterministic policy for this problem is ensured.
We try to acquire such an optimal policy by solving the optimality equations (\ref{eq7}) in the next section.

\section{Solving the Infinite State SMDP Problem}\label{section:procedure}
Iteration-based algorithms, such as value iteration and policy iteration, are widely used to solve the optimality equations for most common discrete-time, finite-state and discounted Markov decision process (MDP) problems.
However, the standard procedure is not readily applied to our problem, which is a continuous-time SMDP with an infinite state space and a long-term average objective.

We solve this problem in three steps.
In Section~\ref{subsec:approximate}, we approximate the infinite state space by a finite state space through ``tail" state aggregation.
In Section~\ref{subsec:dtmdp}, we transform the finite state SMDP to an equivalent discrete time MDP.
Finally in Section~\ref{subsec:rvi}, we use the relative value iteration (RVI) algorithm to solve the average-cost MDP problem.
\subsection{Finite State Approximation}
\label{subsec:approximate}

The SMDP problem has infinite states in $\mathcal{S}=\{0,1,2,\ldots\}$, and is impractical to be solved by numerical methods.
Hence, we truncate the infinite state space to a \emph{finite} state space $\mathcal{\hat{S}}=\{0,1,\ldots,s_{\max},S_{\mathrm{o}}\}$, which replaces the states larger than $s_{\max}$ by an ``overflow" state $S_{\mathrm{o}}$.
In other words, the ``overflow" state $S_{\mathrm{o}}$ is an aggregation of the ``tail" states $\mathcal{S_{\mathrm{tail}}}=\{s_{\max}+1,s_{\max}+2,\ldots\}$. 
The dimension of the finite state space is $|\mathcal{\hat{S}}|=s_{\max}+2$, where $s_{\max}$ needs to be no less than $B_{\max}$.
The rationale of the state space truncation is that the tail probability, defined as the probability of being in the ``tail" states $\mathcal{S_{\mathrm{tail}}}$, decreases with $s_{\max}$.
When $s_{\max}$ is large enough, the ``tail" states are negligible.

In the truncated model, the action space $\mathcal{A}$, the sojourn time distribution $\Gamma_{s,a}(\cdot)$, and the expected sojourn time $y(s,a)$ are the same as before, while the feasible action space at state $S_{\mathrm{o}}$ is $\mathcal{A}_{S_{\mathrm{o}}} 
\equiv \mathcal{A}$ since $s_{\max} \geq B_{\max}$.
Original transitions to the ``tail" states $\mathcal{S_{\mathrm{tail}}}$ are aggregated to $S_{\mathrm{o}}$, and
we assume the number of requests at $S_{\mathrm{o}}$ is  $s_{\max}$. 
The adapted transition probability $\hat{m}(j|s,a)$ for $\forall j,s \in \mathcal{\hat{S}}, \forall a \in \mathcal{A}_s$ is
\begin{equation}
\begin{aligned}
&\hat{m}(j|S_{\mathrm{o}},a) =\\
&\begin{cases}
p_{j-s_{\max}+a}^{[a]}&  j \geq s_{\max}-a, j \neq S_{\mathrm{o}},a \neq 0\\
1-\sum\limits_{i=0}^{a}p_i^{[a]}
& j = S_{\mathrm{o}},a \neq 0 \\
1& j=S_{\mathrm{o}},a=0
\\
0& \text{ otherwise }
\end{cases},
\\
&\hat{m}(j|s,a) \; (s \neq S_{\mathrm{o}})=\\
&\begin{cases}
p_{j-s+a}^{[a]}& j \geq s-a, j \neq S_{\mathrm{o}},a \neq 0 \\
1-\sum\limits_{i=0}^{s_{\max}-s+a}p_i^{[a]}
&
j = S_{\mathrm{o}},a \neq 0 \\
1& j = s+1,s<s_{\max},a=0
\\
1& j=S_{\mathrm{o}},s = s_{\max},a=0
\\
0& \text{ otherwise }
\end{cases}.
\end{aligned}
\label{eqm}
\end{equation}

The unbounded holding cost induced by the infinite states in the primal problem is also erased by the truncation.
Therefore, we introduce an abstract cost $c_{\mathrm{o}}y(s,a) \; (c_{\mathrm{o}} \geq 0)$ to the ``overflow" state $S_{\mathrm{o}}$, working as an estimation of the difference between the expected holding cost at ``tail" states and the holding cost at $s_{\max}$.
The adapted cost $\hat{c}(s,a)$ is
\begin{equation}
\begin{aligned}
\hat{c}(s,a)= 
\begin{cases}
    c(s_{\max},a)+c_{\mathrm{o}}y(s,a) \quad s = S_{\mathrm{o}} \\
    c(s,a) \quad \quad \quad \quad \; s \neq S_{\mathrm{o}}, s \in \mathcal{\hat{S}}
\end{cases},
\end{aligned}
\forall a \in \mathcal{A}_s.
\label{eq13}
\end{equation}
Since $\rho \in [0,1)$, the optimal policy must stabilize the system.
The abstract cost can also be interpreted as an overflow punishment, which pushes the optimal policy away from causing overflow.
Note that the abstract cost $c_{\mathrm{o}}y(s,a)$ in (\ref{eq13}) is rarely mentioned in the literature, without which the problem can be solved as well, but leading to a larger satisfactory $s_{\max}$ and higher computational complexity in iteration algorithms (which will be discussed in Section~\ref{subsec:complexity}).

Let $\mathcal{\hat{P}} \triangleq \{\mathcal{T},\mathcal{\hat{S}} ,\mathcal{A}_s,\hat{m}(\cdot|s,a), \Gamma_{s,a}(\cdot),\hat{c}(s,a)\}$ denote the finite state SMDP model, and the optimality equations for the finite-state average-cost SMDP problem are
\begin{equation}
\begin{aligned}
\hat{h}(s)=\min _{a \in \mathcal{A}_s}\bigg\{\hat{c}(s, a)-\hat{g}y(s, a)+\sum_{j \in \mathcal{\hat{S}}} \hat{m}(j | s, a) \hat{h}(j)\bigg\},
\end{aligned}
\label{optimality3}
\end{equation}
for $\forall s \in \mathcal{\hat{S}}$.
Denote $\hat{g}_*$ as the optimal average expected cost.

Given a stationary deterministic policy as a function $\hat{\pi}:\mathcal{\hat{S}} \rightarrow \mathcal{A}$, the corresponding state transition matrix is $P_{\hat{\pi}} \in \mathbb{R}^{{|\mathcal{\hat{S}}|}\times{|\mathcal{\hat{S}}|}}$.
Suppose that the Markov chain with $P_{\hat{\pi}}$ has a unique stationary distribution $\bm{\mu}=(\mu_0,\mu_1,\ldots,\mu_{S_{\mathrm{o}}})$, where $\mu_s$ is the stationary probability at state $s \in \mathcal{\hat{S}}$.
Then the average expected cost per unit time is
\begin{equation}
    \hat{g}^{\hat{\pi}}= \frac{\sum_{s \in \mathcal{\hat{S}}} \mu_s \cdot \hat{c}(s,\hat{\pi}(s))}{\sum_{s \in \mathcal{\hat{S}}} \mu_s \cdot y(s,\hat{\pi}(s)) }.
    \label{eq14}
\end{equation}

We establish a criterion for assessing the approximation based on the difference in average cost under stabilizing policies:
Let $\Delta^{\hat{\pi}}$ represent the average expected cost contributed by $S_{\mathrm{o}}$ per unit time under policy $\hat{\pi}$:
\begin{equation}
    \Delta^{\hat{\pi}}= \frac{\mu_{S_{\mathrm{o}}} \cdot \hat{c}(S_{\mathrm{o}},\hat{\pi}(S_{\mathrm{o}}))}{\sum_{s \in \mathcal{\hat{S}}} \mu_s \cdot y(s,\hat{\pi}(s)) }.
\label{eq15}
\end{equation}
It is important to note that for a policy that stabilizes the system, the average cost contributed by the ``tail" states should asymptotically decrease to zero as $s_{\max}$ increases. 
Therefore, given a predefined constant $\delta>0$, if $\Delta^{\hat{\pi}} < \delta$, we consider the approximation acceptable with tolerance $\delta$.
If $\Delta^{\hat{\pi}} \geq \delta$, we conclude that the approximation is not acceptable with tolerance $\delta$, and a larger $s_{\max}$ should be selected.
 

\subsection{Associated Discrete-Time MDP}
\label{subsec:dtmdp}
The finite state continuous-time SMDP is associated with a \emph{discrete-time} MDP through a ``discretization" transformation (see Section 11.4 of \cite{puterman1994markov}).
The time slots are denoted by $\tilde{\mathcal{T}}=\{0,1,2,\ldots\}$.
The state space $\mathcal{\hat{S}}$, the action space $\mathcal{A}$ and the feasible action space $\mathcal{A}_s$ for any $s \in \mathcal{\hat{S}}$ keep unchanged in the transformed model.
The transformed cost $\tilde{c}(s, a)$ and the transformed transition probability $\tilde{m}(j | s, a)$ for $\forall j,s \in \mathcal{\hat{S}}, \forall a \in \mathcal{A}_s$ are
\begin{equation}
\begin{aligned}
    \tilde{c}(s, a) \triangleq& \; \hat{c}(s, a) / y(s, a),\\
    \tilde{m}(j | s, a) \triangleq& 
    \begin{cases}
    \eta \hat{m}(j | s, a) / y(s, a)& j \neq s \\ 
    1+\eta[\hat{m}(s| s, a)-1] / y(s, a)& j=s
    \end{cases},
\end{aligned}
\label{transform}
\end{equation}
where $\eta$ satisfies
\begin{equation}
    0<\eta<y(s, a)/(1-\hat{m}(s|s, a)),
\end{equation}
for all $a \in A_s$ and $s \in \mathcal{\hat{S}}$ for which $\hat{m}(s|s, a)<1$. 

By (\ref{eqy}) and (\ref{eqm}), $\eta$ should satisfy
\begin{equation}
  0<\eta<\min\Bigg\{\frac{1}{\lambda},\min_{a \in \mathcal{A}, a \neq 0}\bigg\{
  \frac{1}{{\mu}^{[a]}(1-p_a^{[a]})}, \frac{1}{{\mu}^{[a]}\sum\limits_{i=0}^{a}p_i^{[a]}}
  \bigg\}\Bigg\}.  
\end{equation}
And from experiments we find that the larger the $\eta$ is, the faster the value-based iteration algorithm converges.

The discrete-time MDP model can be denoted by $\tilde{\mathcal{P}} \triangleq \{\tilde{\mathcal{T}},\mathcal{\hat{S}},\mathcal{A}_s,\tilde{m}(\cdot|s,a),\tilde{c}(s,a)\}$.
The transformation (\ref{transform}) serves to standardize costs to a unit time basis, and then adjust the transition structure to align the long-run average cost of the discrete model $\tilde{\mathcal{P}}$ with that of the SMDP model $\mathcal{\hat{P}}$ (refer to Section 11.5.1 in \cite{puterman1994markov} for additional insights into this conversion).

For the average cost MDP problem with $\tilde{\mathcal{P}}$, the optimality equations are
\begin{equation}
\tilde{h}(s)=\min _{a \in {\mathcal{A}}_s}\bigg\{\tilde{c}(s,a)-\tilde{g} +\sum_{j \in \mathcal{\hat{S}}} \tilde{m}(j|s, a) \tilde{h}(j)\bigg\},
\forall s \in 
\mathcal{\hat{S}}
.
\label{eq12}
\end{equation}
According to Proposition 11.4.5 in \cite{puterman1994markov}, if $(\tilde{g},\tilde{h})$ satisfies the discrete-time optimality equations in (\ref{eq12}), then $(\tilde{g},\eta\tilde{h})$ satisfies (\ref{optimality3}).
Let $\tilde{g}_*$ represent the optimal average expected cost in the MDP problem. 
Then, $\tilde{g}_*=\hat{g}_*$ is equivalent to the optimal average expected cost per unit time in the continuous-time SMDP problem.
Therefore, the optimal stationary policy $\tilde{{\pi}}^*$ for the MDP problem, given by
$$
\tilde{{\pi}}^*(s) \in \underset{a \in \mathcal{A}_s}{\arg \min }
\bigg\{\tilde{c}(s,a)-\tilde{g} +\sum_{j \in \mathcal{\hat{S}}} \tilde{m}(j|s, a) \tilde{h}(j)\bigg\}, 
\forall s \in \mathcal{\hat{S}},
$$
is also optimal for the finite state SMDP problem (in section~\ref{subsec:approximate}).
The existence of a solution to (\ref{eq12}) is established through Theorem 8.4.3 in \cite{puterman1994markov}.





\subsection{Relative Value Iteration}
\label{subsec:rvi}
We utilize the value-based iteration algorithm to solve the optimality equations (\ref{eq12}) of the discrete-time MDP problem.
Specifically, for average-cost MDP problems, the standard value iteration is numerically unstable, so we use the relative value iteration algorithm instead\cite{puterman1994markov}.

Let $\mathcal{V} \triangleq \mathbb{R}^{|\mathcal{\hat{S}}|}$ denote the space of value
functions. 
For any value function $h \in \mathcal{V}$, the exact Bellman operator is $\mathcal{L} \; : \; \mathcal{V} \rightarrow \mathcal{V}$, defined as
\begin{equation}
(\mathcal{L}h)(s) \triangleq \min \limits_{a \in A_s}\bigg\{\tilde{c}(s, a)+\sum \limits_{j \in \mathcal{\hat{S}}} \tilde{m}(j | s, a) h(j)\bigg\}.
\end{equation}
The span of a value
function $h \in \mathcal{V}$ is defined as
\begin{equation}
    \text{span}(h) \triangleq \max \limits_{s \in \mathcal{\hat{S}}} h(s) - \min \limits_{s \in \mathcal{\hat{S}}} h(s).
    \label{span}
\end{equation}
Let $H_i$ and $J_i$ be value functions that iterate with $i$, and we describe the relative value iteration algorithm in Algorithm~\ref{alg1}.

\begin{algorithm}
\renewcommand{\algorithmicrequire}{\textbf{Input:}}
\renewcommand{\algorithmicensure}{\textbf{Output:}}
\caption{Relative Value Iteration (RVI)} 
	\label{alg1}
	\begin{algorithmic}
        \REQUIRE {A small positive number $\epsilon>0$.}
        \ENSURE {A stationary deterministic policy $\tilde{\pi}_{\epsilon}: \mathcal{S} \rightarrow \mathcal{A}$.}
	    \STATE {\textbf{step 1:} Set $i=0$, and $H_i(s)=J_i(s)=0$ for all $s \in \mathcal{\hat{S}}$.
        Choose an arbitrary state $s^* \in \mathcal{\hat{S}}$.}
        \STATE{\textbf{step 2:} For all $s \in \mathcal{\hat{S}}$, compute $J_{i+1} (s)= (\mathcal{L}H_i)(s)$, expressed as
        \begin{equation}
             J_{i+1} (s)= \min \limits_{a \in A_s}\bigg\{\tilde{c}(s, a)+\sum \limits_{j \in \mathcal{\hat{S}}} \tilde{m}(j | s, a) H_{i}(j)\bigg\}.
        \label{eq21}
        \end{equation}}
        
        \STATE{\textbf{step 3:} For all $s \in \mathcal{\hat{S}}$, compute}
        \STATE{\quad \quad \quad  \quad \quad \quad $H_{i+1} (s)=J_{i+1}(s) - J_{i+1}(s^*).$}
        \STATE {\textbf{step 4:} If $\text{span}(H_{i+1}-H_{i})< \epsilon$, then for all $s \in \mathcal{\hat{S}}$ compute}
	    \STATE {\quad \quad $\tilde{\pi}_{\epsilon}(s) \in \underset{a \in \mathcal{A}_s}{\arg \min }
\bigg\{\tilde{c}(s, a)+\sum \limits_{j \in \mathcal{\hat{S}}} \tilde{m}(j | s, a) H_{i}(j)\bigg\}.$}
        \STATE{\quad \quad \; \; Otherwise increment $i$ by 1, and return to step 2.}
    \end{algorithmic}
\end{algorithm}
Note that in each iteration, $H_i$ is the renormalized form of $J_i$ by subtracting a common $J_i(s^*)$ from each $J_i(s)$. 
This helps prevent the divergence of value functions in ordinary value iteration, but it does not affect the minimizing actions or the value of $\text{span}(H_{i+1} - H_i)$.
The termination of iteration is triggered when $\text{span}(H_{i+1} - H_i)$ becomes smaller than a predefined constant $\epsilon > 0$.
According to Proposition 6.6.1 in \cite{puterman1994markov}, the Bellman operator $\mathcal{L}$ is a contraction operator over the span of (\ref{span}).
Therefore, the iteration algorithm is guaranteed to terminate.
Moreover, it can be proven that within Algorithm~\ref{alg1}, the value function $H_i$ asymptotically converges to the optimal value function as $i \rightarrow \infty$.
The resulting policy $\tilde{\pi}_{\epsilon}$ is an $\epsilon$-optimal policy. 
In other words, the average expected cost associated with policy $\tilde{\pi}_{\epsilon}$, denoted as $\tilde{g}_{\tilde{\pi}_{\epsilon}}$, satisfies $\tilde{g}_{\tilde{\pi}_{\epsilon}} - \tilde{g}_* < \epsilon$. Detailed proof for this can be found in Section 8.5.5 of \cite{puterman1994markov}.
The computational complexity of Algorithm ~\ref{alg1} is discussed as follows.
Suppose the total number of iterations is $n$. 
It is important to note that $s_{\max} \geq B_{\max}$, and in most cases, $s_{\max}$ is significantly larger than $B_{\max}$.
Also, please note that for $s \geq B_{\max}$, the feasible action space can be represented as $\mathcal{A}_s \equiv \mathcal{A} \equiv \{0\} \cup \mathcal{B}$.
Now, we break down the computational complexity:
The number of multiplications per iteration is approximately $\sum \limits_{s \in \mathcal{\hat{S}}} |\mathcal{A}_s||\mathcal{\hat{S}}| \approx B_{\max}s_{\max}^2$.
The number of additions per iteration is approximately $ \sum \limits_{s \in \mathcal{\hat{S}}} |\mathcal{A}_s||\mathcal{\hat{S}}|+ |\mathcal{\hat{S}}|\approx B_{\max}s_{\max}^2$.
As for space complexity, it is primarily determined by $\tilde{c}(s, a)$ and $\tilde{m}(j | s, a)$.
The storage required for $\tilde{c}(s, a)$ contributes to a space complexity of approximately $\sum \limits_{s \in \mathcal{\hat{S}}} |\mathcal{A}_s| \approx B_{\max}s_{\max}$.
Referring to (\ref{eqm}), the storage needed for $\tilde{m}(j | s, a)$ simplifies to the storage of $p_i^{[a]}$, resulting in a space complexity of approximately $\sum \limits_{a \in \mathcal{B}}(s_{\max}+a+1) \approx B_{\max}s_{\max}$.
Consequently, the overall space complexity is $\mathcal{O}(B_{\max}s_{\max})$, and the time complexity is $\mathcal{O}(nB_{\max}s^2_{\max})$.

It should be noted that the state space of the computed policy $\tilde{\pi}_{\epsilon}$ is $\mathcal{\hat{S}}$, but the ultimate goal is to derive a policy that maps from the infinite state space $\mathcal{S}$ to the action space $\mathcal{A}$.
Therefore, given the policy $\tilde{\pi}_{\epsilon} :  \mathcal{\hat{S}} \rightarrow \mathcal{A}$, we can define its corresponding policy in the original infinite state space as  ${\pi}_{\epsilon}\; : \; \mathcal{S} \rightarrow \mathcal{A}$, using the following equation:
\begin{equation}
\pi_{\epsilon}(s) \triangleq
    \begin{cases}
   \tilde{\pi}_{\epsilon}(s)& \quad  s \leq s_{\max}
    \\
    \tilde{\pi}_{\epsilon}(s_{\max})&
    \quad  s > s_{\max}
    \end{cases}.
\end{equation}
Here, the actions for ``tail" states $\mathcal{S_{\mathrm{tail}}}=\{s_{\max}+1,s_{\max}+2,\ldots\}$ are assigned the same action as for the state $s_{\max}$.
In summary, the RVI algorithm guarantees the derivation of a stationary deterministic policy $\tilde{\pi}_{\epsilon}$, which is $\epsilon$-optimal for the discrete-time MDP problem discussed in Section~\ref{subsec:dtmdp}.
Additionally, thanks to the benefits of the ``discretization" transformation, $\tilde{\pi}_{\epsilon}$ maintains its $\epsilon$-optimality for the finite state SMDP problem introduced in Section~\ref{subsec:approximate}.
When considering the performance of ${\pi}_{\epsilon}$ in the original infinite state SMDP problem (as presented in Section~\ref{sec:formulation}), it is closely tied to the impact of finite state approximation. 
On one hand, with a larger $s_{\max},$ the approximation to the original infinite state SMDP becomes more accurate, enhancing the performance of ${\pi}_{\epsilon}$ in the original problem. 
On the other hand, a larger $s_{\max}$ increases the computational complexity of RVI.
Therefore, as defined in Section~\ref{subsec:approximate}, we can set a tolerance value $\delta$ for approximation. 
Choosing an $s_{\max}$ as small as possible is preferred in terms of complexity, as long as the resulting $\tilde{\pi}_{\epsilon}$ satisfies $\Delta^{\tilde{\pi}_{\epsilon}} < \delta$.
\section{Optimal Policy in Special Cases}\label{sec:special cases}
In the previous section, we introduced a general approach to address the formulated SMDP problem. 
However, in existing literature, specific properties of optimal policies have been discussed for certain special cases. 
Research studies\cite{deb1973optimal,aalto2000optimal} have shown that in scenarios with size-independent service times, optimal policies exhibit a threshold-based structure known as a Q-policy or control limit policy, given certain assumptions.

The concept of the control limit policy is explained as follows:
\begin{definition}
Define a stationary deterministic policy $\pi^{Q}  :  \mathcal{S} \rightarrow \mathcal{A}$ as 
\begin{equation}
    \pi^{Q}(s) =
    \begin{cases}
   0& \quad  s < Q
    \\
    \min(s,B_{\max})&
    \quad  s \geq Q
    \end{cases},
\end{equation}
with a parameter $Q \in \mathbb{N}_{+}$. 
Under such a policy $\pi^{Q}$, a batch service of $\min(s,B_{\max})$ requests will be initiated if and only if the number of awaiting requests at a review point, $s$, exceeds the threshold $Q$.

This policy is called a \textbf{Q-policy}\cite{weiss1979computation}, or a \textbf{control limit policy}\cite{ignall1974optimal} with the threshold $Q$ defined as the \textbf{control limit}.
\label{definition2}
\end{definition}

The necessary assumptions for the specific cases discussed in this section are listed as follows:
\begin{assumption}
Service times are independent and identically distributed (i.i.d.), irrespective of the batch size.
In other words, $l(b) \equiv l$ with $l>0$ independent of $b$, and ${\mu}^{[b]} \equiv \mu \equiv \frac{1}{l}$ is independent of $b$ as well.
\label{ass1}
\end{assumption}
\begin{assumption}
The minimum batch size is $1$, i.e., $B_{\min}=1$.
\label{ass2}
\end{assumption}
\begin{assumption}
    The energy consumed in a batch service is a linear function of $b$, i.e., $\zeta(b)=\beta b+\zeta_0$, with $\beta>0$ and $\zeta_0 \geq 0$.
    \label{ass3}
\end{assumption}

The conclusion regarding the structure of the optimal policy for the specific scenario is as follows.
\begin{proposition}
    If Assumptions 1-3 hold, then there exists a positive integer $Q$, $1 \leq Q \leq B_{\max}$, such that the associated control limit policy ${\pi}^Q$ is an average expected optimal policy for the SMDP model $\mathcal{P}$.
    \label{proposition3}
\end{proposition}

\begin{IEEEproof}
See Appendix~C.
\end{IEEEproof}

\begin{assumption}
    Service times follow an exponential distribution, i.e., $G_b(x) = 1-e^{- {\mu}^{[b]} x}$.
    \label{ass4}
\end{assumption}

In a more special case with exponential service time, the optimal $Q$ value can be computed in the following way: 

\begin{proposition}[Refer to Section 6 in \cite{deb1973optimal}]
    Assume that Assumptions 1-4 hold.
    Combining Assumption 1 and Assumption 4, we have $G_b(x) \equiv G(x) \equiv 1-e^{- \mu x}$, with mean $1/{\mu} \equiv l$. 
    Let $\psi = \lambda / (\lambda + \mu)$, and let $\xi(0<\xi<1)$ denote the unique solution of
    $$(1-\psi)\xi^{B_{\max}+1}-\xi+\psi=0.$$
    Let $\chi=\lambda/\mu$, $r=\xi/(1-\xi)$, and
    \begin{equation}
        D_q = q\{\frac{1}{2}(q+1)+\chi-r\}-r^2\xi^q+r(r-\chi)-\frac{w_2\zeta_0\lambda^2}{w_1}.
        \label{eq30}
    \end{equation}
Then, the optimal $Q$ is the smallest positive integer $q \leq B_{\max}$ for which $D_q \geq 0$.
If there is no positive $q \leq B_{\max}$ such that $D_q \geq 0$, the optimal $Q$ is $B_{\max}$.
\label{proposition4}
\end{proposition}

\begin{collary}
Assuming that Assumptions 1-4 are satisfied, if either $w_2=0$ or $\zeta_0=0$, the optimal value of $Q$ is only influenced by $\chi=\lambda/\mu$ and $B_{\max}$.  
\label{collary1}
\end{collary}
\begin{IEEEproof}
See Appendix~D.
\end{IEEEproof}
It is important to note that the general case under Assumptions 1-3 is intractable, which means that the optimal threshold $Q$ cannot be obtained through explicit computation.
In such cases, a linear search approach can be employed to assess the policy performance of various potential $Q$ values, thereby allowing us to identify the optimal threshold $Q$.

\begin{figure*}
\centering
\includegraphics[width=1\linewidth]{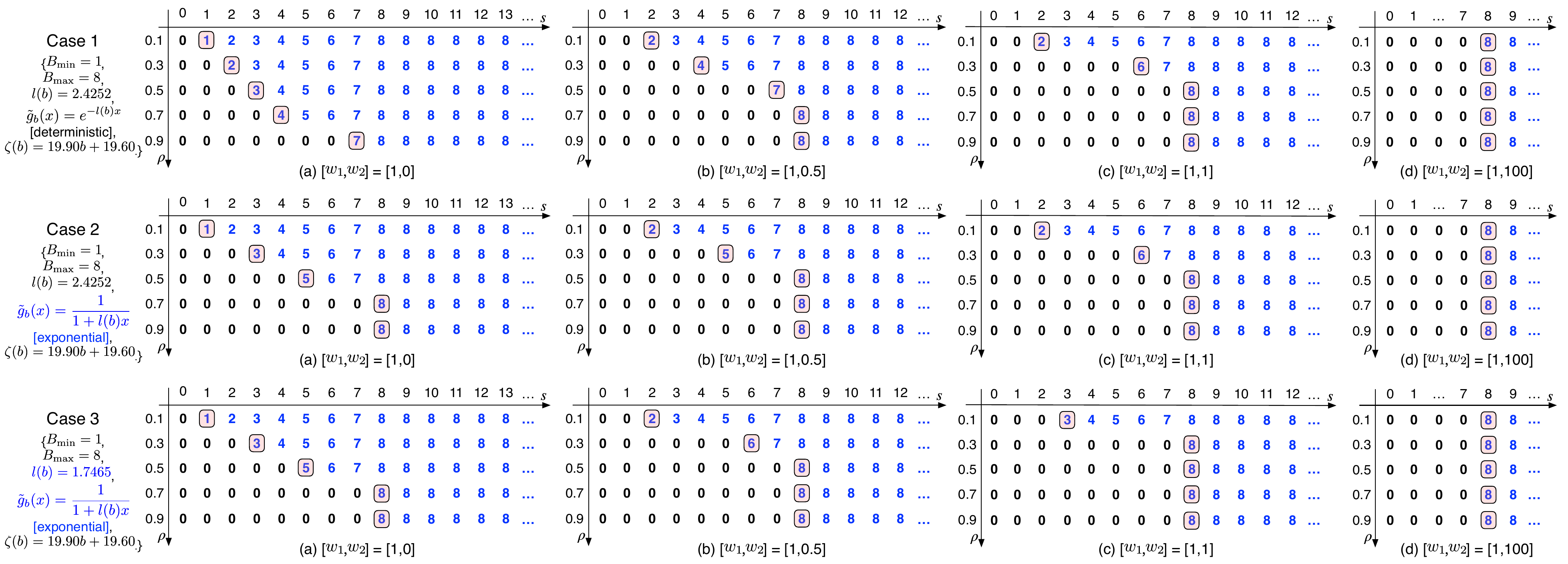}
\caption{The converged SMDP solutions under various parameter settings. The maximum batch size is chosen as $B_{\max}=8$. The weights are (a) $[w_1,w_2]=[1,0]$, (b) $[w_1,w_2]=[1,0.5]$, (c) $[w_1,w_2]=[1,1]$ and (d) $[w_1,w_2]=[1,100]$. The normalized traffic intensity $\rho$ varies in $\{0.1,0.3,0.5,0.7,0.9\}$. All the solutions exhibit a
control limit structure, with the control limits highlighted
by pink boxes.} 
\label{policy_view}
\end{figure*}


\section{Numerical Results}\label{sec:results}
In numerical experiments, we take the GoogLeNet inference on a TESLA P4 as the basic scenario\cite{NVIDIA}.
As depicted in Fig.~\ref{fig:twosub}, the latency and energy functions, which were fitted from the empirical data\cite{NVIDIA}, are $l(b)=0.3051b+1.0524$ ms and $\zeta(b)=19.899b+19.603$ mJ, respectively.
Since the processing time for such image recognition tasks is almost deterministic\cite{ali2020batch}, the Laplace transform of the service time is $\tilde{g}_b(x)=e^{-l(b)x}$. 
The minimum batch size $B_{\min}$ for the ML inference task is $1$, and the maximum batch size $B_{\max}$ is set to $32$ by default.
This basic scenario with deterministic processing times and linear latency and energy functions is acknowledged as a representative case for ML inference serving\cite{ali2020batch,inoue2021queueing,yao2022eais}.


We conduct experiments under varying values of $\rho$ and $w_2$.
It is important to note that $\rho \in (0,1)$ represents the ``normalized" traffic intensity, calculated as the ratio of the absolute traffic intensity (arrival rate) $\lambda$ to the maximum service rate $B_{\max}\mu^{[B_{\max}]}$.
This ratio serves as a measure of the system load.
The weight $w_2$ reflects the importance of power consumption in the overall objective, with $w_1$ fixed at $1$.

\subsection{SMDP Solution}
In this subsection, we visualize the SMDP solutions under various parameter sets, as illustrated in Fig.~\ref{policy_view}.
We construct three scenarios with processing times independent of the batch size, named Cases 1-3, based on the basic scenario.
The maximum batch size $B_{\max}$ is set as 8, for convenience of visualization.
The depicted solutions are the converged results (which remain consistent with increased $s_{\max}$) obtained using the procedure in Section~\ref{section:procedure}.
Each horizontal block in Fig.~\ref{policy_view} corresponds to one scenario, and solutions are obtained under different $\rho$ and $w_2$.
The charts of policies are placed from left to right for increasing $w_2$.
Each row in the chart is a stationary deterministic policy under a certain $\rho$, where each element denotes the action taken at the state corresponding to the column.

From Fig.\ref{policy_view}, it can be observed that for Cases 1-3, where Assumptions 1-3 hold true, all SMDP solutions exhibit a control limit structure, with the control limits highlighted by pink boxes.
This finding concurs with the conclusions in Proposition~\ref{proposition3}. 
Under control limit policies, the system does not serve until the state exceeds a threshold, known as a ``control limit".
Once the state exceeds this control limit, the system serves a maximum available batch of requests. 
It can be seen that the control limit increases with $w_2$. 
When $w_2$ is as large as $100$, the control limits under different traffic intensities are all $B_{\max}$. 
This is reasonable because the importance of power consumption grows with $w_2$, and the energy is better saved with a larger batch size.

For Case 2 and Case 3 that satisfy Assumptions 1-4, the optimal control limits can be explicitly calculated using Proposition~\ref{proposition4}.
We observe that the control limits $Q$ of the obtained SMDP solutions are in alignment with the directly calculated results, which also validates the effectiveness of the proposed general solving procedure.
It is further observed that when $[w_1,w_2]=[1,0]$, the SMDP solutions in Case 3 are exactly the same as those in Case 2. 
This can be explained by Collary~\ref{collary1}: when $w_2=0$, the control limits are solely influenced by $B_{\max}$ and $\chi=\rho B_{\max}$, which means that they are only influenced by values $B_{\max}$ and $\rho$.
Moreover, when $w_2 \neq 0$, it can be seen that the control limits in Case 3 are equal to or larger than those in Case 2.
Note that the batch service rate in Case 2 is $\theta(b)=\frac{b}{2.4252}$, while in Case 3 it is $\theta(b)=\frac{b}{1.7465}$.
As a result, Case 3 offers a greater marginal benefit from increasing the batch size compared to Case 2,  leading to its control limits no less than those of Case 2. 

Furthermore, upon examining the solutions in a broader set of cases (see Appendix~E), we have observed that in more general situations with characteristics such as size-dependent batch service time, a minimum batch size greater than 1, or a nonlinear energy consumption function, the control limit structure may not be applicable or maintained.

\subsection{Performance Comparisons}\label{subsec:comp}




In this subsection, we compare the performance of the obtained SMDP solutions with other benchmark batching policies.  
The benchmark policies encompass the greedy batching policy, as well as static batching policies with batch sizes of $b=8,16,32$. 
Under the greedy batching policy, the server processes the largest feasible batch of current requests.
In static batching policies, the server consistently processes batches in a fixed batch size and waits for new incoming requests if there are insufficient requests. 
The static batching policy with $b=32$ represents a special case known as the maximum batching policy in this context, where $B_{\max}=32$.

In what follows, we first showcase that the SMDP-based policies always yield the lowest average cost compared to other benchmark policies. 
Then, we present the two-dimensional figures illustrating latency and energy measurements, highlighting the superior performance of SMDP solutions from a Pareto perspective.

Furthermore, we demonstrate that the SMDP-based policy can enhance the satisfaction of delay requirements, as it produces a lighter-tailed distribution.

\subsubsection{Overall Objective}
\begin{figure}
    \centering    
    \includegraphics[width=0.75\linewidth]{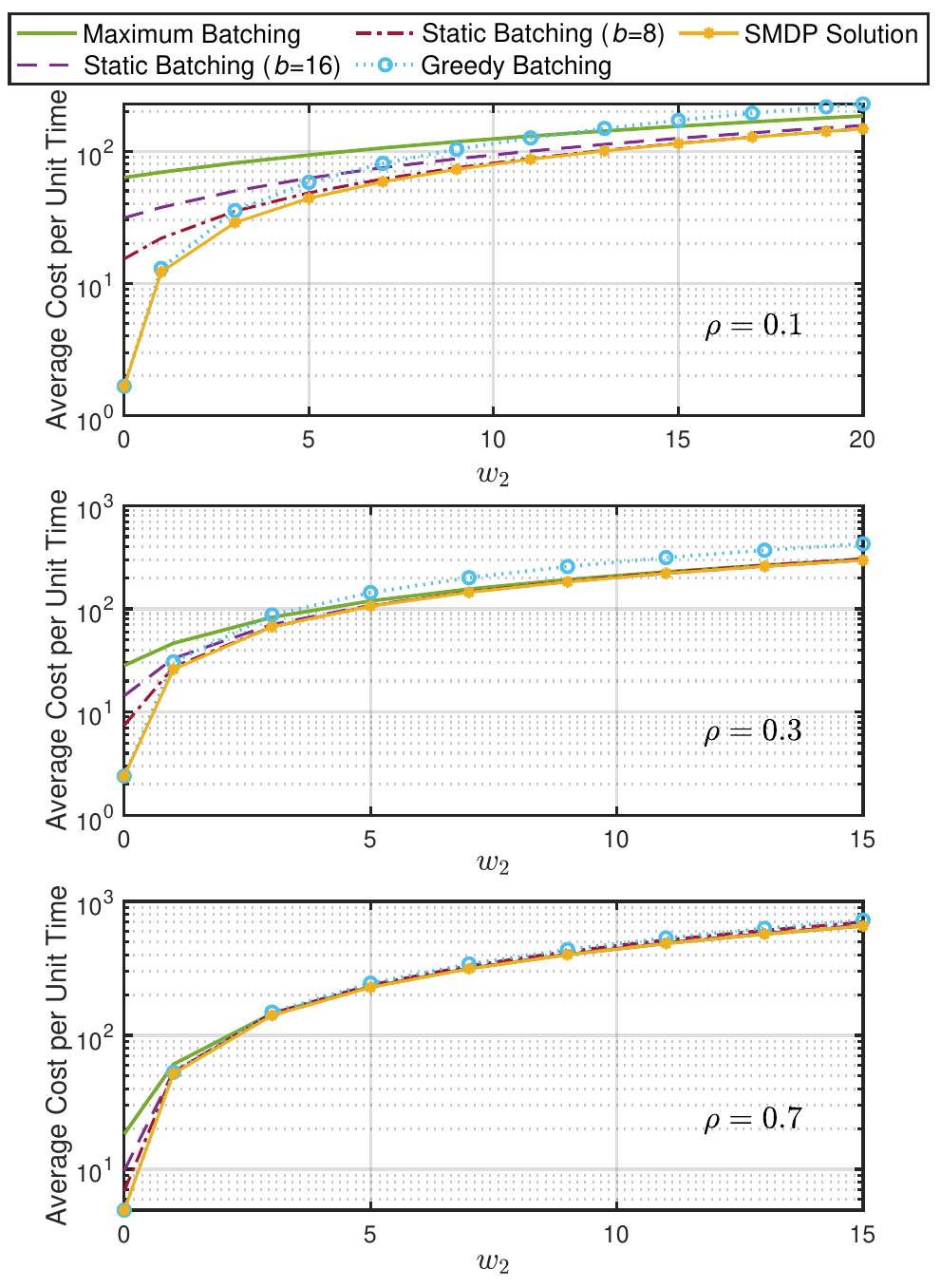}
    \vspace{-3mm}
    \caption{Comparison of different policies on the average cost per unit time under $\rho=0.1, 0.3, 0.7$, with $w_1=1$ and $w_2$ ranging from $0$ to $15$.}
    \label{fig:data_3_log}
\end{figure}

We first compare the overall objective, namely the average cost per unit time, under three levels of traffic intensities, $\rho=0.1,0.3$ and $0.7$, as shown in Fig.~\ref{fig:data_3_log}. 
The weight for the latency term, $w_1$, is fixed at $1$, while the weight for the power consumption term, $w_2$, ranges from $0$ to $15$.
The objectives are computed using Eq.(\ref{eq14}).
It is observed that the SMDP solutions always achieve the lowest (best) average cost per unit time among all policies under various parameter settings.

When $w_2$ is close to zero, the objective primarily focuses on \emph{latency}.
In such cases, we observe that the cost of the greedy batching policy is close to that of the SMDP-based policy.
Meanwhile, the costs associated with static batching policies are higher and increase with the batch size $b=8,16,32$, with the maximum batching policy incurring the highest (worst) cost. 
This suggests that the latency introduced by serving with larger batches is comparable or even greater than the latency saved by increased batch service rate.
When $w_2$ reaches a large value, the objective is primarily influenced by \emph{power consumption}.
In such cases, it is observed that the maximum batching policy yields nearly the lowest cost, approaching that of the SMDP solution.
(Unfortunately, for $\rho=0.1$, a value of $w_2=15$ is not large enough to observe this phenomenon.)
This observation is consistent with the results in Fig.~\ref{policy_view}.
Meanwhile, in such cases, the greedy batching policy works poorly and incurs a much higher cost than other policies, due to its limited parallelism.
In most common scenarios, the weighted average cost is not dominated by a single term, and the other two static batching policies ($b=8,16$) can achieve a proper balance between latency and energy, thus approaching the SMDP solution under certain parameter scales.

The comparison of the overall objective has two main limitations:
(1) The overall objective lacks a unified metric and is not sufficiently informative, as it combines power consumption and latency through a weighted sum.
(2) The value of the overall objective can become infinitely large as $w_2$ increases, resulting in a wide range that is difficult to visualize, especially under high load conditions.
Therefore, we will analyze the objective factors separately in what follows.


\subsubsection{Objective Pairs}
\begin{figure*}
    \centering
    \subfloat[The $(\overline{W},\overline{P})$ pairs for SMDP solutions.]{
    \includegraphics[width=0.26\linewidth]{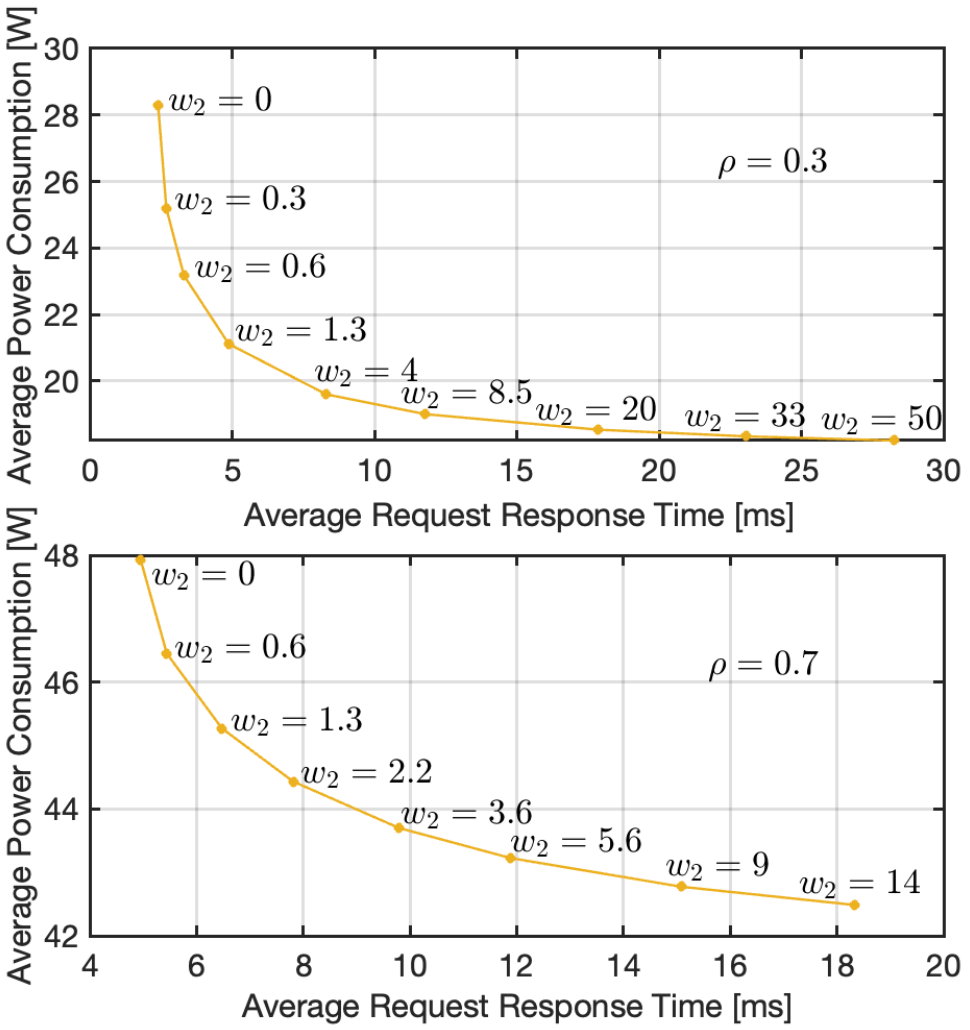}
    \label{fig:data_3_tradeoff}
}
\hspace{0mm}
    \subfloat[The latency-power consumption tradeoff.]{%
        \includegraphics[width=0.33\textwidth]{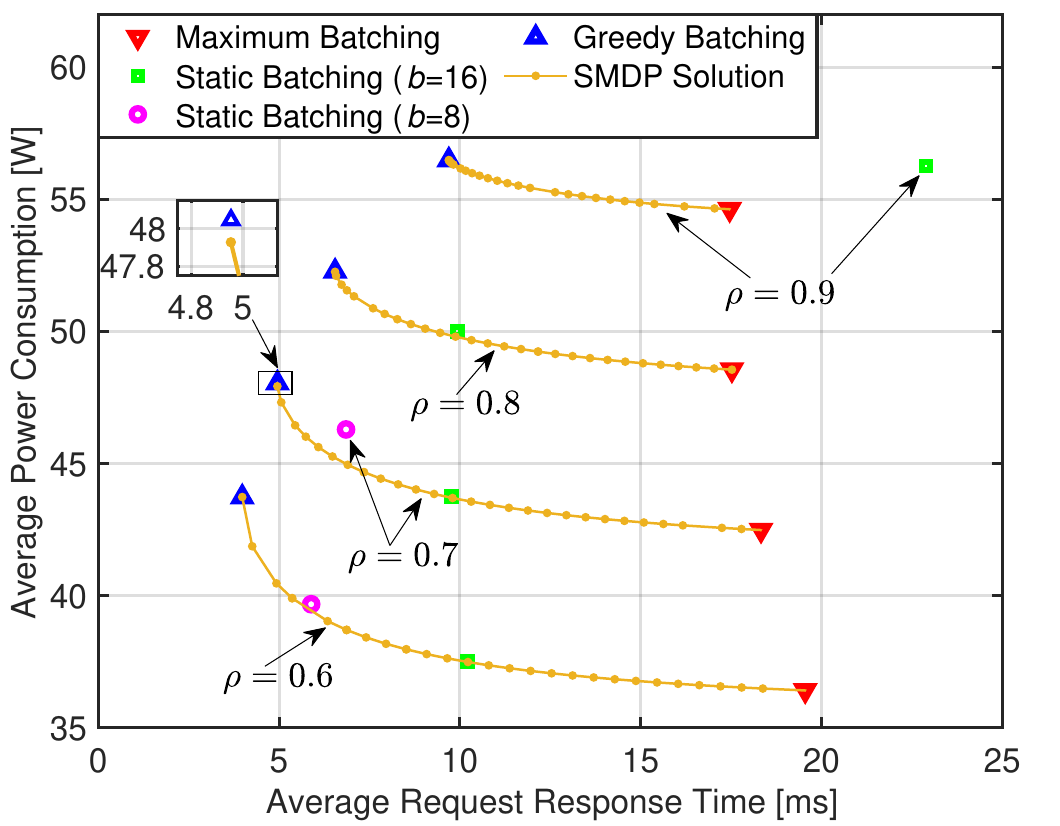}%
        \label{fig:case6_tradeoff}%
    }
    \hspace{0mm} 
    \subfloat[The latency-energy efficiency tradeoff.]{%
        \includegraphics[width=0.33\textwidth]{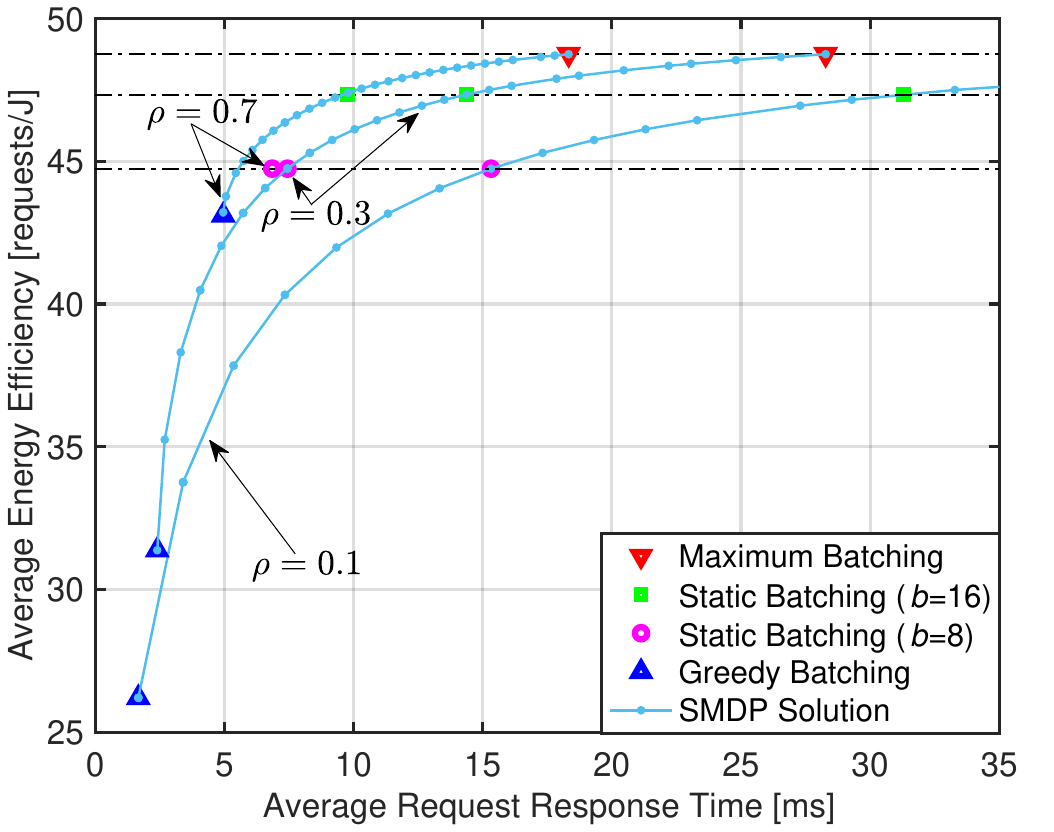}%
        \label{fig:case6_tradeoff_ee}%
    }
    \caption{The latency-energy tradeoff curves for different policies under various load conditions.}
    \label{fig:case6}
\end{figure*}


We now focus on the two factors in the multi-objective optimization (as formulated in Eq.(\ref{eq:multi-obj})):
(a) Average request response time, or long-term average latency, denoted as $\overline{W}$;
(b) Average power consumption, denoted as $\overline{P}$, calculated by dividing the long-term energy consumption by the time period. 
The unit of average power consumption is measured in Watt (W).
Additionally, we introduce average energy efficiency, representing the average number of requests processed per Joule of energy (calculated by ${\lambda}/{\overline{P}}$), as an alternative measure of power consumption. 



By fixing $w_1=1$ and varying $w_2$, different SMDP solutions can be obtained through the proposed scheme.
Accordingly, a set of $(\overline{W},\overline{P})$ pairs is acquired.
The latency-power consumption tradeoff curves for these  $(\overline{W},\overline{P})$ pairs under $\rho=0.3$ and $\rho=0.7$ are illustrated in  Fig.~\ref{fig:case6}\subref{fig:data_3_tradeoff}.
It can be seen that as the weight for power consumption, $w_2$, increases, the average power consumption decreases while the average request response time increases, thus forming the latency-energy tradeoff.
After acquiring this curve, an appropriate weight can be selected according to the requirements. For example, if the average request response time is required to be less than 5 ms when 
$\rho=0.3$, the maximum weight whose corresponding SMDP solution meets this requirement should be selected, which is $w_2=1.3$ in this case. By selecting the weight and batching policy in this manner, the least power consumption is achieved while satisfying the latency requirement. The method for choosing an appropriate $w_2$
under a power constraint is similar.


Fig.~\ref{fig:case6}\subref{fig:case6_tradeoff} illustrates the $(\overline{W},\overline{P})$ pairs of different policies under various load conditions.
Furthermore, Fig.~\ref{fig:case6}\subref{fig:case6_tradeoff_ee} depicts the interplay between latency and energy efficiency. 
By varying the values of $(w_1,w_2)$, the solutions obtained from the weighted-objective SMDP demonstrate a flexible balance between latency and energy, forming the tradeoff curves.
In contrast, the objective pairs of benchmark policies are represented as separate points that remain unchanged with the weights. 
Moreover, it can be seen from Fig.~\ref{fig:case6}\subref{fig:case6_tradeoff} (or Fig.~\ref{fig:case6}\subref{fig:case6_tradeoff_ee}) that the SMDP objective pairs are positioned to the lower (or upper) left  of those of the benchmark policies, indicating that the SMDP solutions consistently outperform other benchmark policies in a Pareto-optimal sense.

The latency-power consumption pairs associated with the maximum batching policy precisely correspond to the right endpoints of the SMDP's tradeoff curves. 
This observation aligns with our findings in Fig.~\ref{policy_view} and Fig.~\ref{fig:data_3_log} under conditions where $w_2$ is large.
This is reasonable, given that the maximum batching policy exhibits the highest energy efficiency among all policies. 
There is no alternative policy that can achieve equal or lower power consumption with a smaller latency.
The latency-power consumption pairs associated with the greedy batching policy are situated near the left endpoints of SMDP's tradeoff curves. 
It is essential to highlight that, although not evident, the greedy batching policy is at a slight disadvantage compared to the SMDP solution.
For instance, when $\rho=0.7$, magnifying the plot around the left corner of the SMDP's tradeoff reveals that the objective pair $(\overline{W},{\overline{P}})$ of the greedy batching policy (the blue triangle) is in the upper right relative to the objective pair of an SMDP solution (an orange dot).

Several points of static batching policies with $b=8$ and $b=16$ are close to the SMDP's tradeoff curve, indicating that these policies can effectively approximate SMDP solutions under specific parameters. 
This observation aligns with the findings in Fig.~\ref{fig:data_3_log}.
However, in certain cases, the superiority of the SMDP-based policy over static batching becomes evident. 
For example, when $\rho=0.7$, the latency-power consumption (or energy efficiency)  pair of static batching with $b=8$ is positioned above (or below) the SMDP curve, implying higher power consumption (or lower energy efficiency) compared to an SMDP-based policy with equal latency. 
Furthermore, the static batching policy with $b=8$ fails to stabilize the system when $\rho \geq 0.8$.
Similarly, when $\rho=0.9$, static batching with $b=16$ results in significantly longer latency compared to the SMDP solutions.

\subsubsection{Latency Distribution and Percentile Analysis}
\begin{figure*}
    \centering
    \subfloat[The empirical CDFs of request response time under different batching policies.]{
    \includegraphics[width=0.29\linewidth]{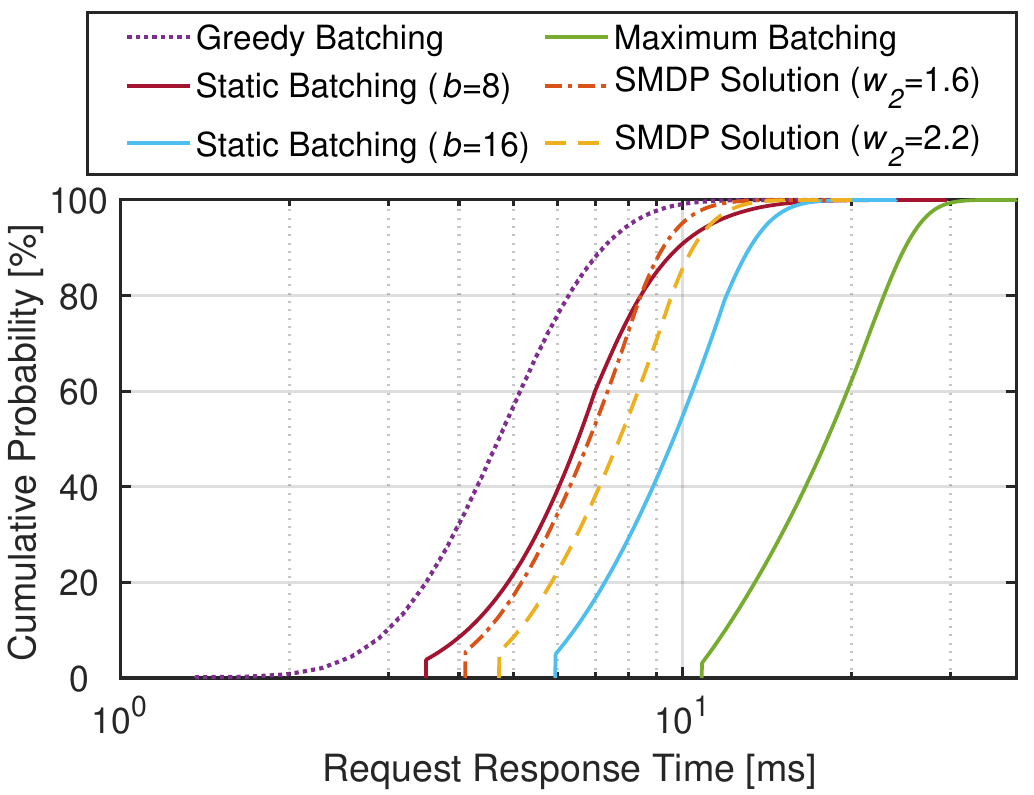}
    \label{cdf_case6_rho7}
}
\hspace{2mm}
    \subfloat[The tradeoff between the latency at the 95th percentile and the average power consumption.]{%
        \includegraphics[width=0.30\textwidth]{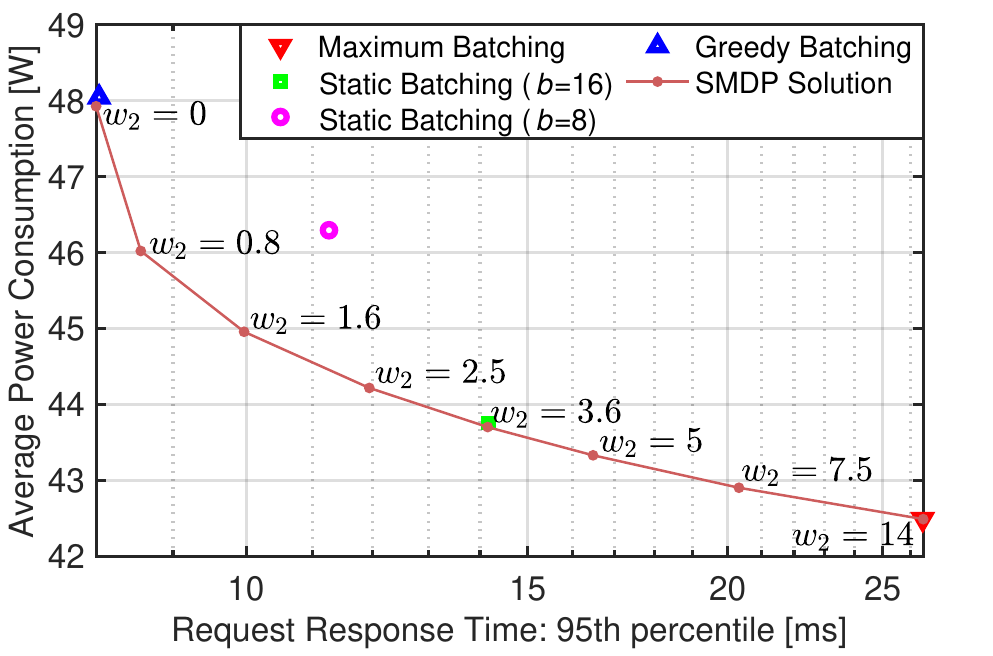}%
        \label{L95_case6_rho7}%
    }
    \hspace{2mm} 
    \subfloat[The tradeoff between the latency satisfaction percentage and the average power consumption.]{%
        \includegraphics[width=0.30\textwidth]{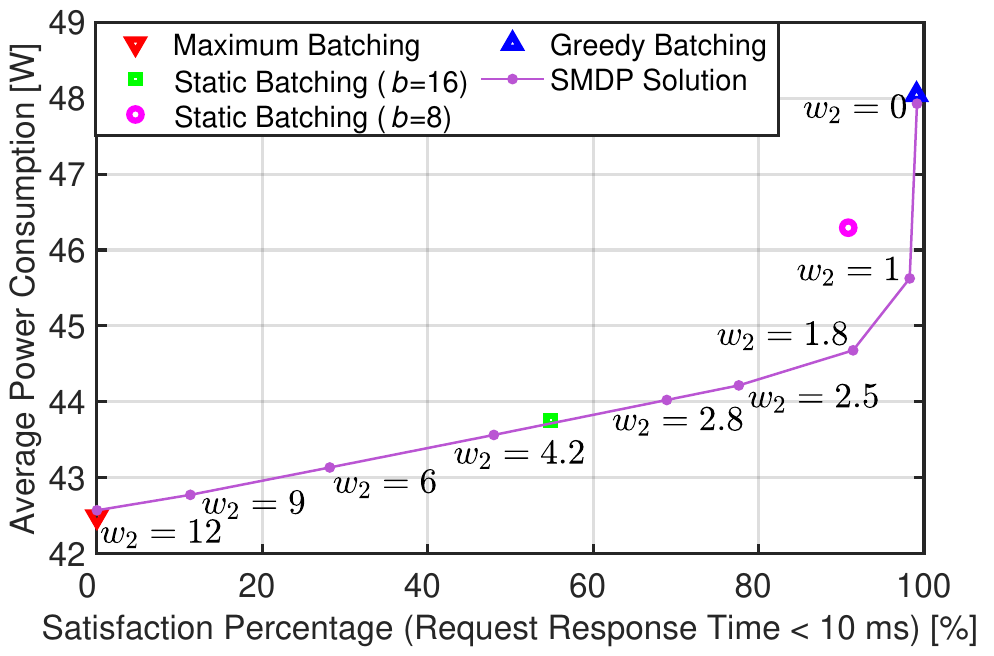}%
        \label{Lless10_case6_rho7}%
    }
    
    \caption{The latency distribution and percentile analysis under different policies, with $\rho=0.7$ and $w_1$ fixed at $1$.}
    \label{fig:case6_r7}
\end{figure*}


In real applications, the average request response time studied in our formulated framework may not be the primary concern. 
Instead, the service level objective (SLO) usually specifies that the request response time at a certain percentile must meet some latency bound.
Therefore, we conduct simulations and subsequently analyze the distribution and percentiles of latency. 

Fig.~\ref{fig:case6_r7}\subref{cdf_case6_rho7} demonstrates the empirical CDFs of request response time for different batching policies under a load condition of $\rho=0.7$, with each CDF based on $1.66 \times 10^6$ latency data points.
A CDF positioned further to the left represents a better policy, as it achieves lower latency for a higher proportion of requests. 
Therefore, the latency performance of the benchmark policies, from best to worst, is ranked as follows: greedy batching, static batching with $b=8$,  static batching with $b=16$, and maximum batching.
The CDFs of the selected SMDP solutions with $w_2=1.6$ and $w_2=2.2$ intersect the CDF of static batching with $b=8$ with an upward crossing at the intersection points.
This indicates that the CDFs of these SMDP solutions are lighter-tailed, and at percentiles beyond the intersection points, the request response times are shorter than that of static batching with $b=8$.

\begin{table}[h]
\scriptsize
\caption{Comparison of average power consumption and request response times at different percentiles under $\rho=0.7$.}
    \centering
    \begin{tabularx}{0.45\textwidth}
    {m{2.5cm}<{\centering}
m{1.2cm}<{\centering}
m{1.25cm}<{\centering} m{1.25cm}<{\centering}}
       \hline
       Policy  & Static Batching ($b=8$) & SMDP Solution ($w_2=1.6$) & SMDP Solution ($w_2=2.2$)\\
       \hline
       \vspace{0.2mm}
       $\overline{P}$ [W] & $46.27$ & $\bm{44.96}$ & $\bm{44.41}$\\
       $\overline{W}$ [ms] & $6.85$ & $6.90$ & $7.81$\\
       $W$: 50th percentile [ms] & $6.51$ & $6.83$ & $7.72$\\
       $W$: 90th percentile [ms] & $9.85$ & $\bm{9.23}$ & $10.45$\\
       $W$: 95th percentile [ms] & $11.34$ & $\bm{9.96}$ & $\bm{11.24}$\\
       \hline
    \end{tabularx}
    \label{table_cdf_statistics}
\end{table}

Table~\ref{table_cdf_statistics} presents more comprehensive data including average power consumption, average request response time, and request response times at the 50th, 90th, and 95th percentiles for static batching with 
$b=8$, as well as SMDP solutions with $w_2=1.6$ and $w_2=2.2$. 
While both SMDP solutions achieve lower power consumption, they result in longer average response times compared to static batching with $b=8$. 
However, the SMDP solution with $w_2=1.6$ improves the request response times at the 90th and 95th percentiles compared to static batching with $b=8$.
Similarly, the SMDP solution with $w_2=2.2$ provides a shorter request response time at the 95th percentile compared to static batching with $b=8$.



Suppose the SLO specifies that the 95th percentile of request response time  must be less than 10 ms.
Then, we can either obtain and plot the data pairs of $($latency at the 95th percentile$,\overline{P})$ under various weights, as illustrated in Fig.~\ref{fig:case6_r7}\subref{L95_case6_rho7}, or plot the data pairs of $($satisfaction percentage for the 10 ms constraint$,\overline{P})$, as shown in Fig.~\ref{fig:case6_r7}\subref{Lless10_case6_rho7}.
Both exhibit tradeoff trends similar to those in Fig.~\ref{fig:case6}\subref{fig:data_3_tradeoff}.
Therefore, the maximum (or minimum) weight $w_2$ should be selected such that the corresponding SMDP solution results in a 95th percentile request response time of less than 10 ms (or a satisfaction percentage for the 10 ms constraint greater than 95\%).
This ensures that the SLO constraint is met while minimizing the power consumption.

\subsection{Performance Comparisons in Other Settings}\label{subsec:impact}
All experiments in Section~\ref{subsec:comp} are conducted in the default configuration.
Therefore, in this subsection, we study some other typical cases and demonstrate the comparison results in these settings.
\subsubsection{Stronger Batching Effect in Batch Service Rate}


\begin{figure*}
    \centering
    \subfloat[Setting: batch-size independent service time.]{
    \includegraphics[width=0.31\linewidth]{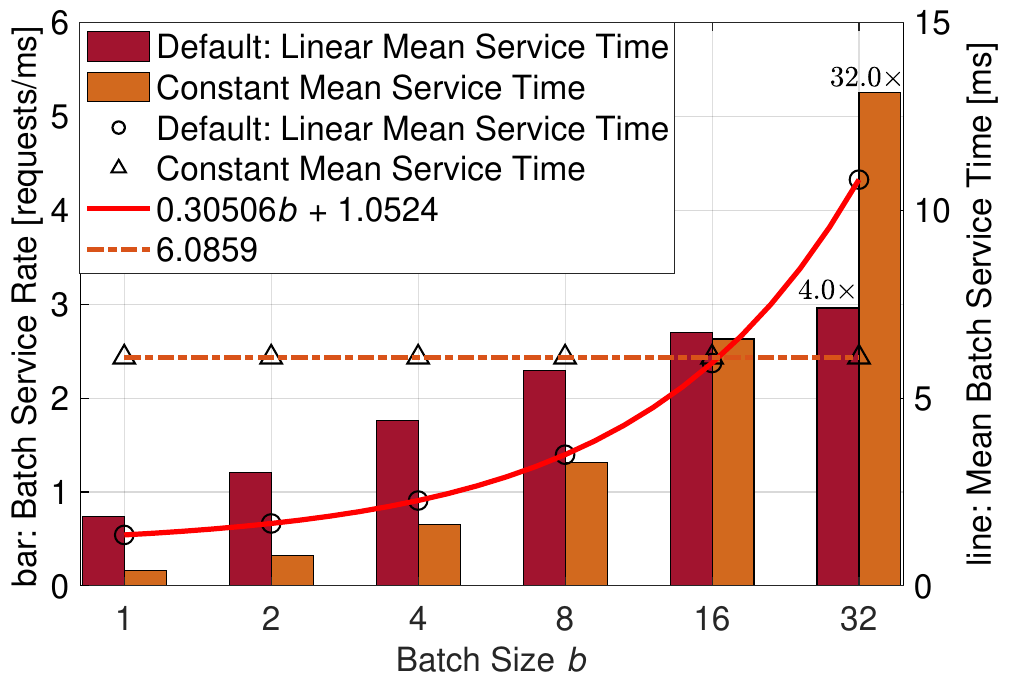}
    \label{case1_setting}
}
\hspace{0mm}
    \subfloat[Result: the latency-power consumption tradeoff.]{%
        \includegraphics[width=0.323\textwidth]{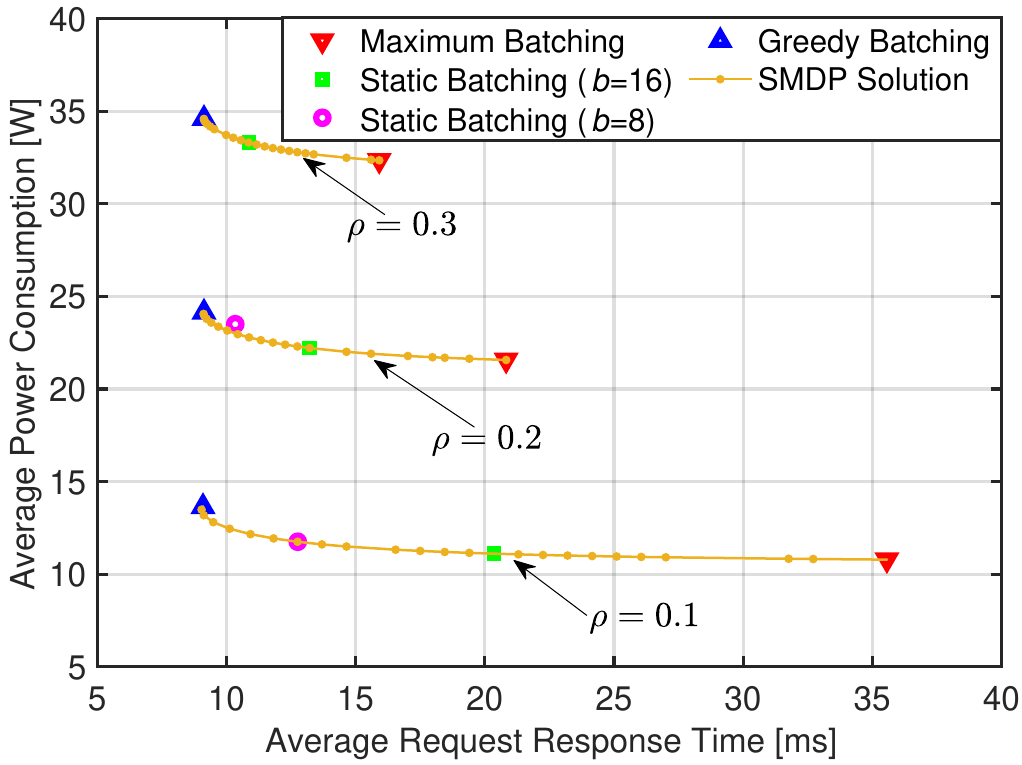}%
        \label{fig:case1_tradeoff}%
    }
    \hspace{0mm} 
    \subfloat[Result: the latency-energy efficiency tradeoff.]{%
        \includegraphics[width=0.32\textwidth]{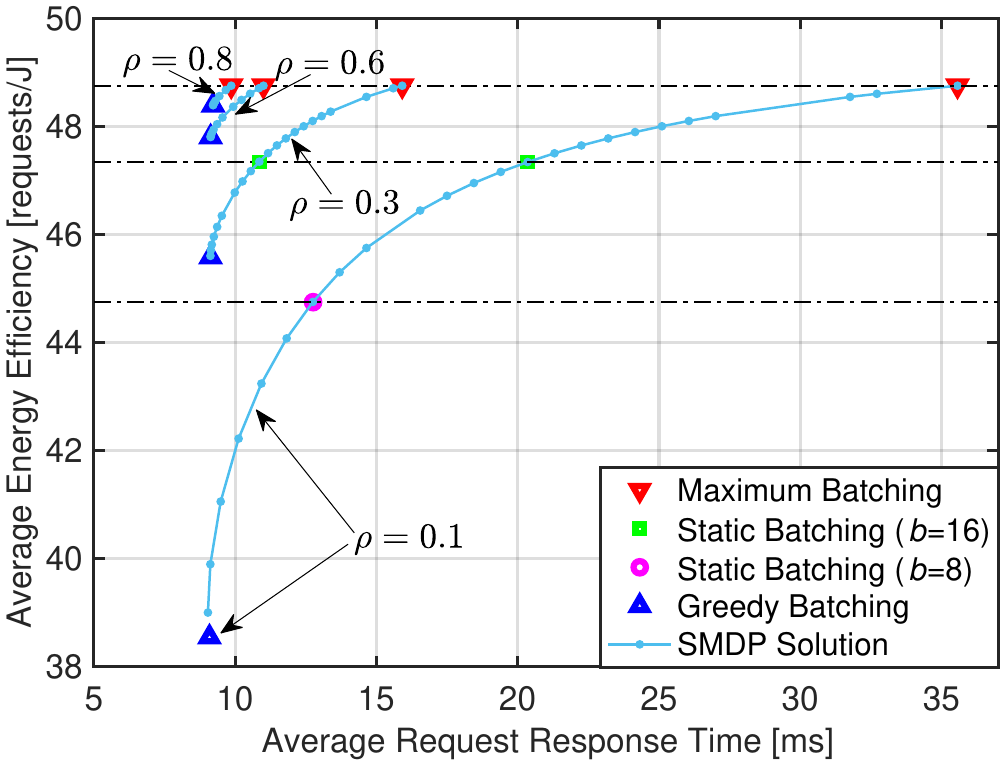}%
        \label{fig:case1_tradeoff_ee}%
    }
    \caption{Illustration and comparison of results with batch-size independent service time.}
    \vspace{-1mm}
    \label{fig:case1}
\end{figure*}
\begin{figure*}
    \centering
    \subfloat[Setting: logarithmic energy consumption function.]{
    \includegraphics[width=0.32\linewidth]{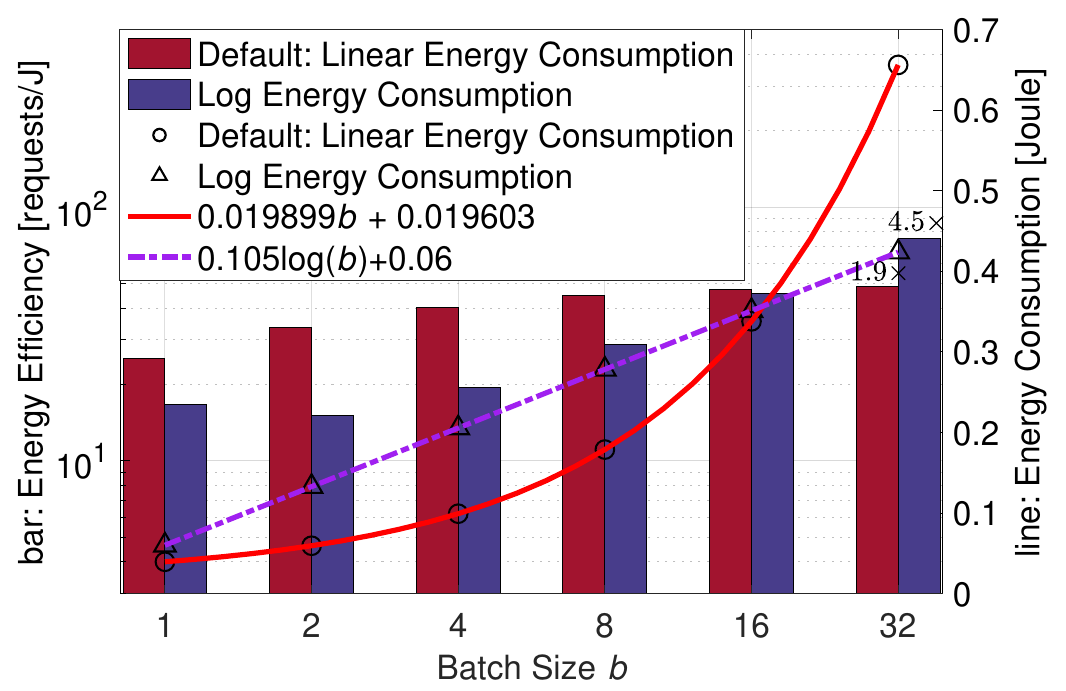}
    \label{case7o_setting}
}
\hspace{0mm}
    \subfloat[Result: the latency-power consumption tradeoff.]{%
        \includegraphics[width=0.32\linewidth]{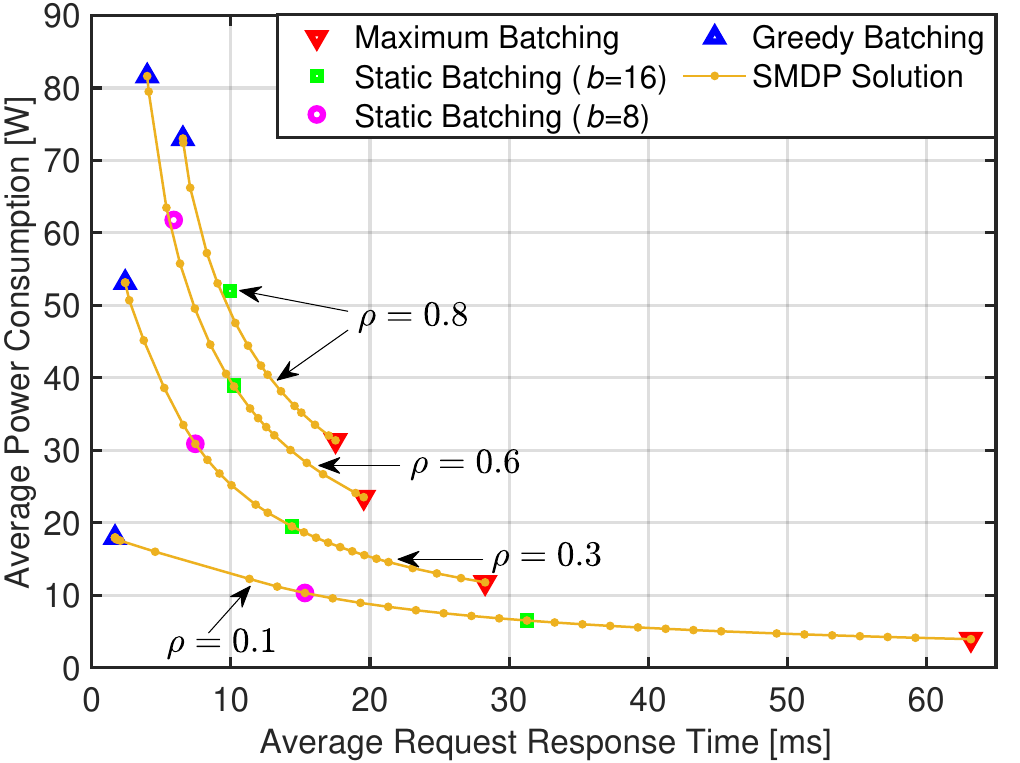}%
        \label{fig:case7o_tradeoff}%
    }
    \hspace{0mm} 
    \subfloat[Result: the latency-energy efficiency tradeoff.]{%
        \includegraphics[width=0.32\linewidth]{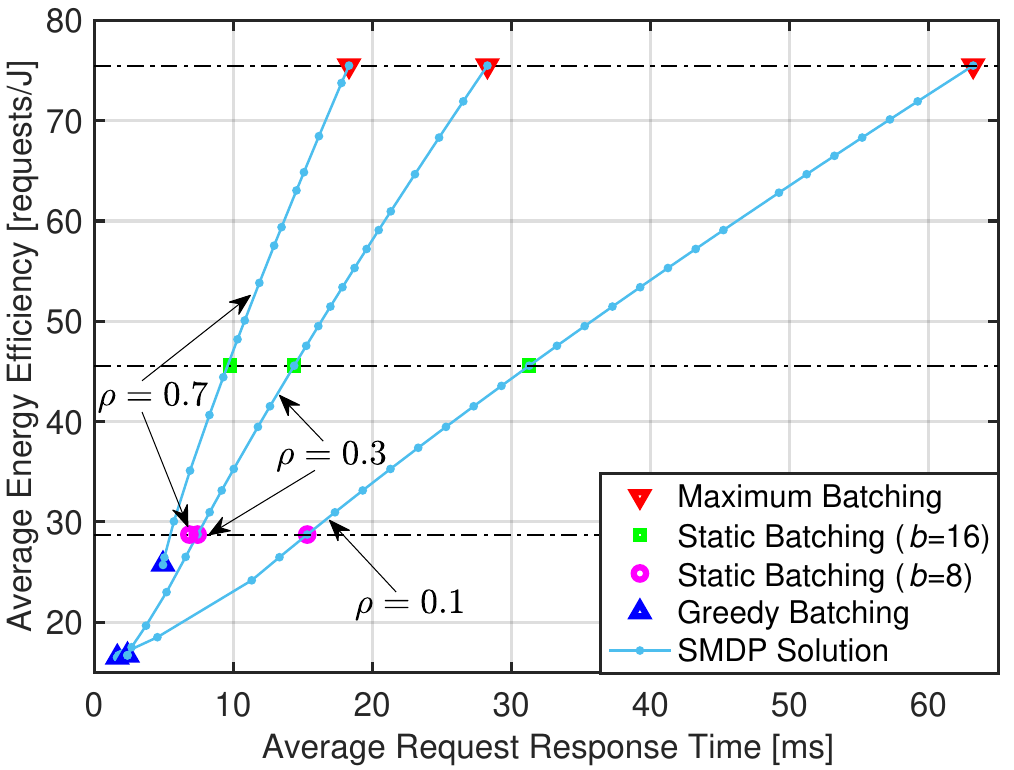}%
        \label{fig:case7o_tradeoff_ee}%
    }
    \caption{Illustration and comparison of results with a logarithmic energy consumption function.}
    \label{fig:case7o}
\end{figure*}

In this part, we modify the mean batch service time function of the default scenario to a constant value of $l(b)=6.0859$ ms, representing ideal parallelism.
Scenarios involving running inference models on powerful processors can approximate this ideal batching, such as the inference of InceptionV2 with float16 on a Titan V with batch sizes up to fifty\cite{hanhirova2018latency}.
As shown in Fig.~\ref{fig:case1}\subref{case1_setting}, whereas the default batch service rate increases sub-linearly with the batch size $b$, the constant service time leads to a linear increase in the batch service rate with $b$.
Consequently, at $b=32$, the batch service rate achieves a speedup of 32 times compared to service without batching ($b=1$), whereas it is only 4 times in the default setting.

The latency-power consumption pairs for different policies under various load conditions are illustrated in Fig.~\ref{fig:case1}\subref{fig:case1_tradeoff}, and the latency-energy efficiency pairs are shown in Fig.~\ref{fig:case1}\subref{fig:case1_tradeoff_ee}.
It is observed that the SMDP-based policies consistently outperform other benchmarks.
Additionally, a few notable observations for this special case include:
(1) The latency of the greedy batching policy shows minor growth with increasing load, compared to the significant growth in the default setting (see Fig.~\ref{fig:case6}\subref{fig:case6_tradeoff}). 
The reason is that, in the default setting, the experienced service time grows with the average batch size as $\rho$ increases, and the service rate does not increase as quickly as with constant service time, worsening the situation.
(2) The latency of the maximum batching policy is much lower than that in the default case and decreases rapidly with increasing $\rho$.
This is because the batch service rate of maximum batching is significantly higher than that in the default case, and an increased traffic load reduces the waiting time for forming a batch.
(3) When $\rho$ is 0.6 or higher, the latency of the maximum batching policy is very close to that of the work-conserving policy.
This follows the previous observation, as a sufficiently high load can effectively activate the power of the maximum batching policy.
This insight suggests that under high load conditions, if the latency of maximum batching is acceptable, there is no need to compute or select an SMDP policy.

\subsubsection{Stronger Batching Effect in Energy Efficiency}

In this part, we modify the energy consumption function of the default scenario to $\zeta(b)=105 \log(b)+ 60$ mJ, which is a logarithmic function of $b$.
As illustrated in Fig.~\ref{fig:case7o}\subref{case7o_setting}, the default energy consumption function $\zeta(b)$ increases linearly with the batch size $b$, resulting in energy efficiency that grows sub-linearly with $b$.
In contrast, with the logarithmic energy function, the energy efficiency increases super-linearly with the batch size $b$.
Notably, the energy efficiency continues to improve significantly after $b$ exceeds 8, while in the default setting, it remains relatively stable.

The latency-power consumption pairs for different policies under various load conditions are illustrated in Fig.~\ref{fig:case7o}\subref{fig:case7o_tradeoff}, and the latency-energy efficiency pairs are demonstrated in Fig.~\ref{fig:case7o}\subref{fig:case7o_tradeoff_ee}.
It is observed that the SMDP-based policies consistently perform as well as or better than other benchmarks. Notable observations for this case include:
(1) The power consumption of the work-conserving policy decreases as $\rho$ increases from 0.6 to 0.8, unlike in the default setting where power consumption rises with higher load. 
This is due to the significant increase in energy efficiency with larger batch sizes.
(2) The latency-power consumption tradeoff curve is much steeper compared to the default setting. 
This is because the wider power consumption range and consistent latency range result in a larger absolute value of the tradeoff derivative.
Furthermore, compared to the default setting, the energy efficiency increases more rapidly at relatively large latency values.
Therefore, a small increase in latency can lead to substantial power savings in this scenario.

\subsubsection{Impact of the Distribution of Service Time}
In the default setting, the service time follows a deterministic distribution with a coefficient of variation (CoV) of 0.
However, in scenarios such as inference running with interference from other tasks, the service time is typically stochastic, and a larger CoV represents more complex and severe interference.
Therefore, in this part, we conduct experiments with three additional types of service time distributions while keeping the same $l(b)$. 
(a) Erlang distribution with a Laplace transform given by $\tilde{g}_b(x)={(\frac{1}{1+0.5l(b)x})}^2$, which has a CoV of 0.5.
(b) Exponential distribution with a Laplace transform given by $\tilde{g}_b(x)=\frac{1}{1+l(b)x}$, which has a CoV of 1.
(c) Hyperexponential distribution with a Laplace transform given by $\tilde{g}_b(x)=\frac{2}{3} \times \frac{1}{1+0.5l(b)x} + \frac{1}{3} \times \frac{1}{1+2l(b)x}$, which has a CoV of 2.

\begin{figure*}
    \centering
    \subfloat[The CDFs of service time distributions when $b=8$.]{
    \includegraphics[width=0.334\linewidth]{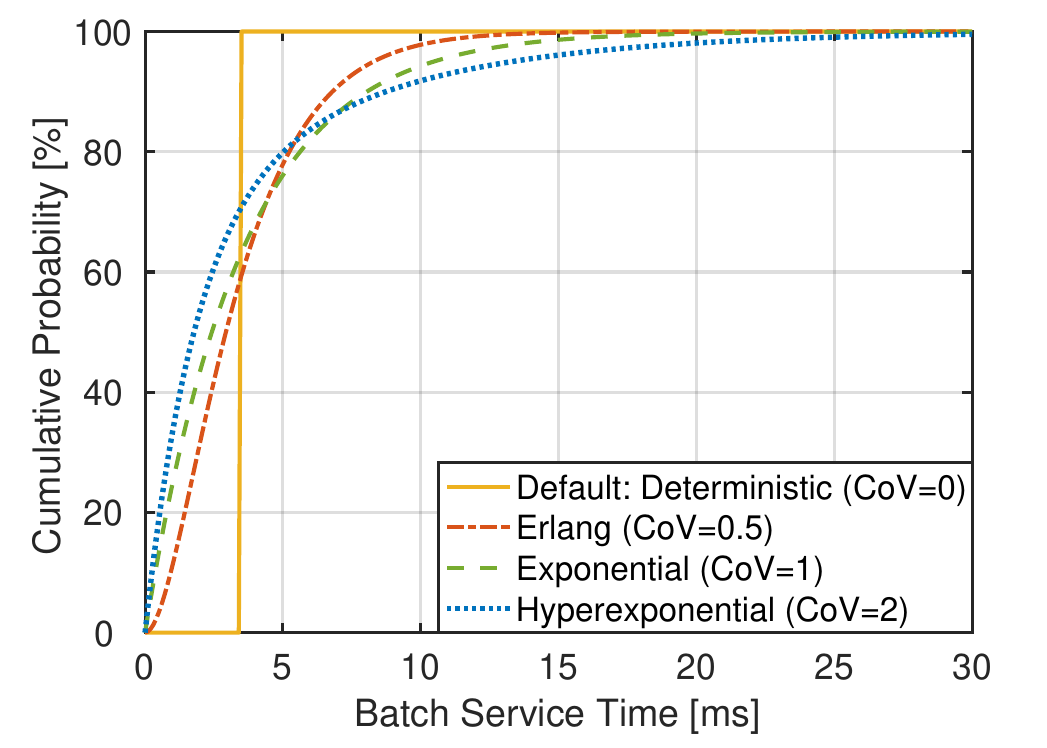}
    \label{cdf_covs}
}
\hspace{-1.5mm}
    \subfloat[The latency-power consumption tradeoff under $\rho=0.3$ and $\rho=0.7$.]{%
        \includegraphics[width=0.63\linewidth]{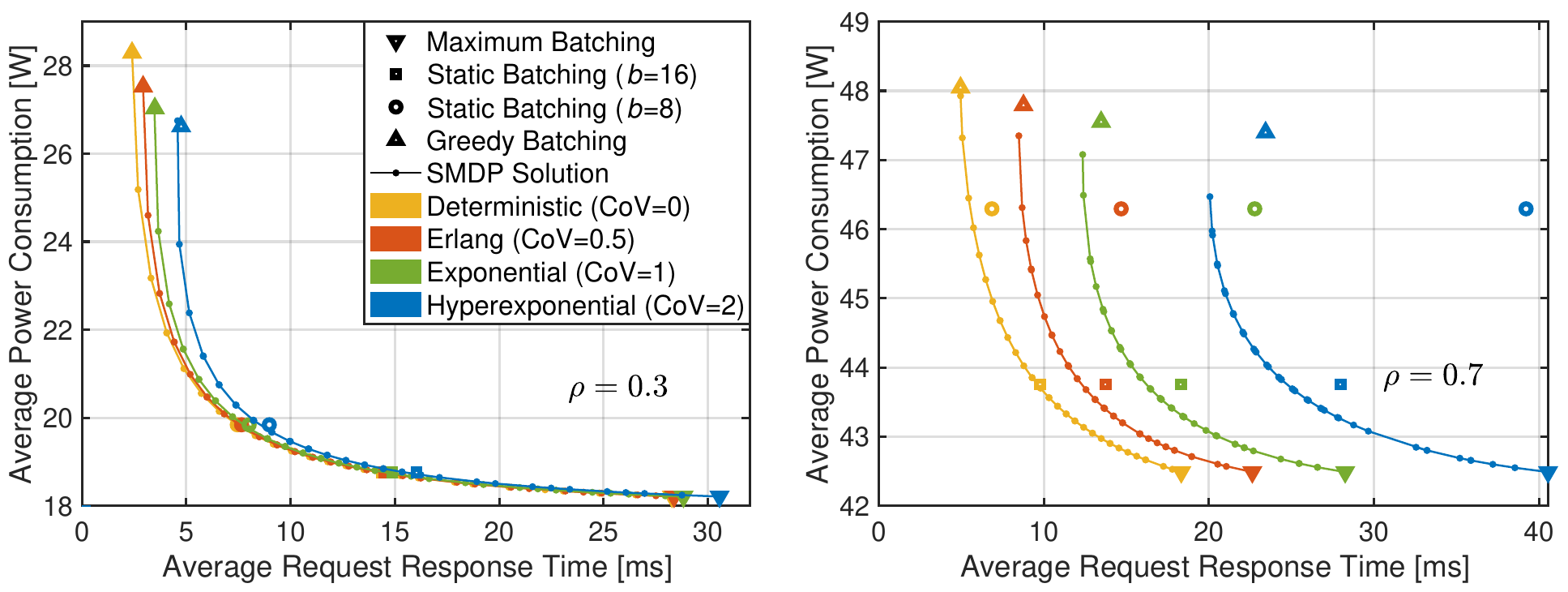}%
        \label{results_covs}%
    }
    \caption{Comparisons of CDFs and latency-power consumption pairs under different service time distributions.}
    \label{fig:covss}
\end{figure*}

The CDFs of these service time distributions for $b=8$ are illustrated in Fig.~\ref{fig:covss}\subref{cdf_covs}.
It can be seen that the tail of the service time distribution becomes heavier as the CoV increases.
Furthermore, as shown in the latency-power consumption curves in Fig.~\ref{fig:covss}\subref{results_covs}, the average latency for a given power consumption increases with the CoV.
This effect is more pronounced under high load conditions (e.g., $\rho=0.7$) compared to low load conditions (e.g., $\rho=0.3$).
This observation aligns with our expectations, as the average latency increases with the CoV due to its corresponding increase in the second moment, as shown in Eq.(\ref{cost_step}).
Additionally, the average power consumption under the greedy batching policy decreases with increasing CoV, reflecting the increase in the average batch size.

\subsection{Efficiency of the Solving Procedure}\label{subsec:complexity}

\begin{table*}[h!]
\scriptsize
\caption{Evaluation of approximations acceptable with tolerance $\delta=0.001$ under different $c_{\mathrm{o}}$.}
    \centering
    \begin{tabular}{cccccc}
       \hline
       $c_{\mathrm{o}}$  & $10000$ & $1000$ & $100$ & $10$ & $0$ \\
       \hline
       $\min s_{\max}$ & $89$ & $78$ & $\bm{70}$ & $161$ & $192$ \\
       $\text{Iterations}$ & $1847$ & $1635$ & $\bm{1483}$ & $10000$ & $10000$ \\
       $\text{Space Complexity}$ & $2848$ & $2496$ & $\bm{2240}$ & $5152$ & $6144$ \\
       $\text{Time Complexity}$ & $4.68 \times 10^{8}$ & $3.18 \times 10^{8}$ & $\bm{2.33 \times 10^{8}}$ & $8.29 \times 10^{9}$ & $1.18 \times 10^{10}$ \\
       $\Delta^{\pi}$ & $9.36 \times 10^{-4}$ & $9.78 \times 10^{-4}$ & $8.36 \times 10^{-4}$ & $6.14 \times 10^{-12}$ & $1.30 \times 10^{-14}$ \\
       $\hat{g}^{{\pi}}$ & $66.1384$ & $66.1383$ & $66.1377$ & $66.1374$ & $66.1374$  \\
       \hline
    \end{tabular}
    \label{table}
\end{table*}

\begin{figure}
\centering
\includegraphics[width=0.70\linewidth]{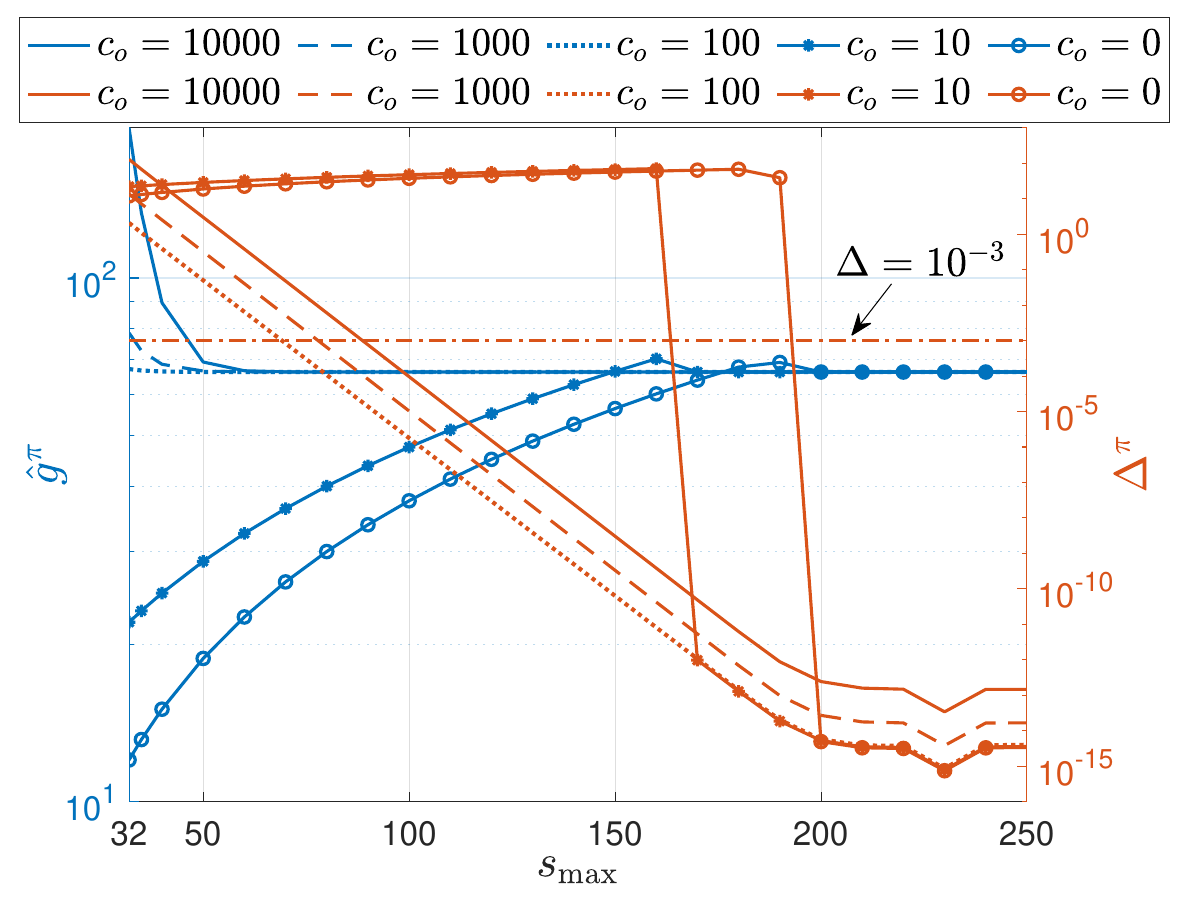}
\vspace{-3mm}
\caption{The evolution of $\hat{g}^{{\pi}}$ (the average cost per unit time) and $\Delta^{\pi}$ (the average cost contributed by $S_{\mathrm{o}}$ per unit time) regarding $s_{\max}$ under different $c_{\mathrm{o}}$, where $s_{\max}$ and $c_{\mathrm{o}}$ are the parameters in finite state approximation.}
\label{convergence}
\end{figure}

We want to evaluate the accuracy and complexity under different abstract costs in the finite state approximation.
As mentioned in Section~\ref{subsec:approximate}, there are two parameters in the approximation: $s_{\max}$ and $c_{\mathrm{o}}$, which determine the dimension of the state space and the abstract cost, respectively.
Given a specific approximate model with certain $s_{\max}$ and $c_{\mathrm{o}}$, a policy ${\pi}_{(c_{\mathrm{o}},s_{\max})}$ is calculated following the procedure outlined in Sections~\ref{subsec:dtmdp} and \ref{subsec:rvi}, and it serves as an approximation to the optimal policy.
The corresponding average cost per unit time $\hat{g}^{{\pi}_{(c_{\mathrm{o}},s_{\max})}}$, evaluated in the state space $\mathcal{\hat{S}}$, can be obtained by Eq.(\ref{eq14}).
The accuracy of the approximation is assessed using $\Delta^{{\pi}_{(c_{\mathrm{o}},s_{\max})}}$, as detailed in Eq.(\ref{eq15}), representing the average cost contributed by $S_{\mathrm{o}}$ per unit time.
For simplicity, we use $\hat{g}^{{\pi}}$ and $\Delta^{\pi}$ to denote these metrics in the following text.
We conduct experiments in the basic scenario, with $\rho=0.9$ and $[w_1,w_2]=[1,1]$.
The stopping parameter in RVI is set to $\epsilon=0.01$.
Additionally, we impose a maximum iteration value in the RVI process as $\rm{iter}_{\max}=10000$.

In Fig.~\ref{convergence}, we illustrate the evolution of $\hat{g}^{{\pi}}$ and $\Delta^{\pi}$ with $s_{\max}$ ranging from 32 to 250.
The $\hat{g}^{{\pi}}$ with $c_{\mathrm{o}}=10000, 1000, 100$ decreases and converges around $s_{\max}=35, 50, 70$, while $\hat{g}^{{\pi}}$ with $c_{\mathrm{o}}=10, 0$ increases and converges around $s_{\max}=170, 200$.
We can infer that the abstract cost with $c_{\mathrm{o}}=10000, 1000, 100$ (or $10,0$) overestimates (or underestimates) the impact of ``tail" states, leading to the $\hat{g}^{{\pi}}$ mostly larger (or smaller) than the convergence value.
From the orange curves, we observe that $\Delta^{\pi}$ decreases with $s_{\max}$, and almost converges when $s_{\max}$ exceeds 200.
A sharp drop is observed for $\Delta^{\pi}$ under $c_{\mathrm{o}}=10, 0$. 
This is due to the underestimated impact of ``tail" states with $c_{\mathrm{o}}=10, 0$, resulting in a lower estimated value for $a=0$ (``wait") in the RHS of Eq.(\ref{eq21}). 
Consequently, the computed policy is to always wait, until $s_{\max}$ is large enough for the cost of waiting to be comparable to the cost of serving.
Although the converged values of $\Delta^{\pi}$ are no more than $10^{-13}$, we only need an approximation acceptable with tolerance $\delta$, and we choose $\delta=0.001$.
In Table~\ref{table}, we list the minimum values of $s_{\max}$ that satisfy the approximation requirement.
The $\Delta^{\pi}$ and $\hat{g}^{{\pi}}$, the number of RVI iterations, as well as the space and time complexity corresponding to the minimum $s_{\max}$ are also recorded.
It can be seen that all $\Delta^{\pi}$ are less than $0.001$, and the differences among $\hat{g}^{{\pi}}$ are also no greater than $0.001$.
The least required $s_{\max}$ is $70$, observed in the approximation with $c_{\mathrm{o}}=100$.
Compared to the ordinary finite state approximation with $c_{\mathrm{o}}=0$, the minimum $s_{\max}$ decreases from $192$ to $70$.
Consequently, the space complexity reduces by $63.5\%$, and the time complexity decreases by $98\%$.
Furthermore, approximations with $c_{\mathrm{o}}$ larger (or smaller) than $100$ exhibit an increasing trend in complexity due to the growing overestimation (or underestimation).

In addition to the state aggregation method used for finite state approximation in our work, the literature also explores approximate iteration algorithms to implement finite state approximation within iteration algorithms\cite{thomas1985finite,white1982finite}.
A comparison between our proposed scheme and two representative approximate iteration algorithms (detailed in Appendix~F) shows that our approximation procedure outperforms these algorithms in both convergence speed and result accuracy, especially when the abstract cost $c_{\mathrm{o}}$ is included.

\section{Conclusion}\label{sec:conclusion}
In this paper, we have studied the dynamic batching problem for online serving, where the batch service time is dependent on the batch size.
The problem is formulated as an SMDP with the objective of minimizing the weighted sum of average response time and average power consumption. 
The inherent complexities of this SMDP problem, characterized by an infinite state space, an average (non-discounted) objective, and unbounded costs, make it challenging to efficiently solve using traditional methods.
To overcome these challenges, we have introduced a solution procedure consisting of finite state approximation, ``discretization" transformation, and relative value iteration. 
The computational complexity is largely reduced owning to the introduction of an abstract cost.
Then, we have conducted comprehensive numerical experiments across various parameter settings. 
The overall average cost and the tradeoff between average latency and average power consumption are depicted under different parameter setups. 
Comparisons with benchmark batching policies further showcase the superiority of the SMDP solutions.

Compared to many existing dynamic batching schemes, our proposed solution is theoretically derived, rather than relying on repeated trials.
As a result, our scheme can be computed offline, alleviating the system from the burden of additional complex modules.
Despite focusing on average objectives, statistics related to the SLO requirements, such as the satisfaction percentage for a certain latency constraint, can be obtained through offline simulations. 
When running in real time, it then becomes easy to find the most suitable weight and its corresponding batching policy that minimizes power consumption while satisfying the SLO requirement.
For bursty or non-stationary traffic arrival processes, which are common in real systems, they can be approximated as temporal compositions of Poisson process periods. 
Specifically, in the case of MMPPs, they are exact temporal compositions of Poisson process periods.
By detecting phases and applying the proposed method to each period, such traffic can be effectively managed.
We also plan to explore dynamic batching schemes for multiple processors, incorporating both inter- and intra-processor parallelism in future work.

\appendices
\section{Proof of Proposition~1}\label{appA}
To prove the proposition, we first introduce Lemma~\ref{Prop2}, which extends Theorem 3 from \cite{sennott1989average} to our specific context.

\begin{lemma}[Extended from Theorem 3 in \cite{sennott1989average}]
Let $d(s)=\underset{a \in \mathcal{A}_s}{\min}\{d(s,a)\}$.
Assume $\underset{s \rightarrow \infty}{\lim} d(s) = \infty$ and for every $a \in \mathcal{B}$, assume $G_a(0)<1$ and  $\frac{1}{{\mu}^{[a]}}<\infty$.
Assume there exist a service parameter $a$ and nonnegative integer $n$ such that $c(s,a) \leq W(s)$, a nonnegative polynomial of degree $n$, $G_a(t)$ has finite $(n+1)$st moment and satisfies $\lambda \frac{1}{{\mu}^{[a]}} < B_{\max}$. Then there exists an expected average cost optimal stationary deterministic policy.
\label{Prop2}
\end{lemma}

Then, we validate these assumptions within our framework.

In $\mathcal{P}$ we have $\underset{s \rightarrow \infty}{\lim} d(s,0) = \underset{s \rightarrow \infty}{\lim} w_1\frac{s}{{\lambda}} =\infty$, and for $a \neq 0$, $\underset{s \rightarrow \infty}{\lim} d(s,a) = \underset{s \rightarrow \infty}{\lim} w_1(\frac{s}{{\lambda}}+\frac{1}{2}\mathbb{E}[G_a^2]{\mu}^{[a]}) =\infty$.
Thus, $\underset{s \rightarrow \infty}{\lim} d(s) = \underset{s \rightarrow \infty}{\lim} \underset{a \in \mathcal{A}_s}{\min}\{d(s,a)\}=\infty$.
For every $a \in \mathcal{B}$, $G_a(0)<1$ and  $\frac{1}{{\mu}^{[a]}}<\infty$ hold, according to Section~III.

When $a=B_{\max}$, there is $c(s,B_{\max})=w_2{\zeta}(B_{\max}) +{w_1}(\frac{s}{\lambda {\mu}^{[B_{\max}]}}+\frac{1}{2}\mathbb{E}[G^2_{B_{\max}}]) \leq W(s) \triangleq c(s,B_{\max})$, where $W(s)$ is a nonnegative polynomial of degree one.
Furthermore, $G_{B_{\max}}(t)$ is assumed to have finite second moment, and $\lambda \frac{1}{{\mu}^{[B_{\max}]}} < B_{\max}$ also holds (see Section~III).
That is to say, there exist a service parameter $a=B_{\max}$ and nonnegative integer $n=1$ that satisfy the assumptions in Lemma \ref{Prop2}.

Therefore, by Lemma \ref{Prop2}, there exisits an average expected optimal stationary deterministic policy for the SMDP model $\mathcal{P}$.

\section{Proof of Proposition~2}\label{appB}
The state process $\{s_m^{{\pi}}\}_{m=0}^{\infty}$ induced by a stationary deterministic policy ${\pi}$, defined as $s_{m+1}^{{\pi}}=f(s_m^{{\pi}}, {\pi}(s_{m}^{{\pi}}))$ with $f(\cdot)$ representing a stochastic evolution, is a Markov chain. 
To get the equations corresponding to the optimal stationary deterministic policies, we need to first discuss the chain structure of the transition matrices of Markov chains generated by stationary
policies\cite{puterman1994markov}.
\begin{definition}
    The SMDP model is \textbf{unichain} if the Markov chain corresponding to \textbf{every} deterministic stationary policy has a single recurrent class and a possibly empty set of transient states.
\end{definition}

\begin{lemma}
    The SMDP model $\mathcal{P}$ is unichain.
    \label{Prop1}
\end{lemma}

\begin{IEEEproof}
    Note that under \emph{any} policy, each state $s$ can reach its neighboring state $s+1$ in one step, since $m(s+1|s,a)>0 , \forall s \in \mathcal{S}, \forall a \in \mathcal{A}_s$. 
    This is valid because $m(s+1|s,0)=1$ and $m(s+1|s,a)=p_{a+1}^{[a]}>0$, for $a \neq 0$. 
    Since state $s+1$ is accessible from state $s$, we know that if state $s$ is a recurrent state, $s+1$ should also be recurrent.
    Therefore, the Markov chain of some stationary policy never contains more than one closed irreducible recurrent class, which concludes the proof. 
\end{IEEEproof}

Then we are able to obtain the optimality equations.
Since $\mathcal{S}$ is countable and the SMDP model is unichain (by Lemma \ref{Prop1}), the optimality equations for such average cost SMDP problems are obtained according to Theorem 11.4.4 in \cite{puterman1994markov}.

\section{Proof of Proposition~3}\label{appC}
The holding cost function is $v(n)=\frac{w_1}{\lambda} n$.
Then, for each value of $s$, it can be observed that $v(s+1)-v(s)=\frac{w_1}{\lambda}>0$.
Additionally, as stated in Section~III, the assumption holds that $\lambda l(B_{\max}) < B_{\max}$.
Combined with Assumption 1, we have $\lambda l< B_{\max}$.
Consequently, the conditions of Theorem 5.3 in \cite{deb1973optimal} are satisfied.
Thus, applying Theorem 5.3 from \cite{deb1973optimal} completes the proof.

\section{Proof of Collary~1}\label{appD}
In Proposition~4, when $w_2=0$ or $\zeta_0=0$, we observe that $D_q$ is solely affected by $q$, $\chi$, and $\xi$.
    Here, $\xi$ is influenced by $\psi$ and $B_{\max}$.
    It is also worth noting that $\psi = \lambda / (\lambda + \mu) = \chi / (\chi+1)$ is determined by $\chi$.
    As a result, $D_q$ is determined by $q$, $\chi$ and $B_{\max}$, when $w_2=0$ or $\zeta_0=0$.
    Therefore, the optimal $Q$, which is the smallest positive integer $q \leq B_{\max}$ such that $D_q \geq 0$, is only influenced by $\chi=\lambda/\mu$ and $B_{\max}$. 
    It remains unaffected by the absolute values of $\lambda$ and $\mu$, when $\chi$ is given.

\section{SMDP Solution Visualization for Broader Cases}\label{appF}
\begin{figure*}
\centering
\includegraphics[width=1\linewidth]{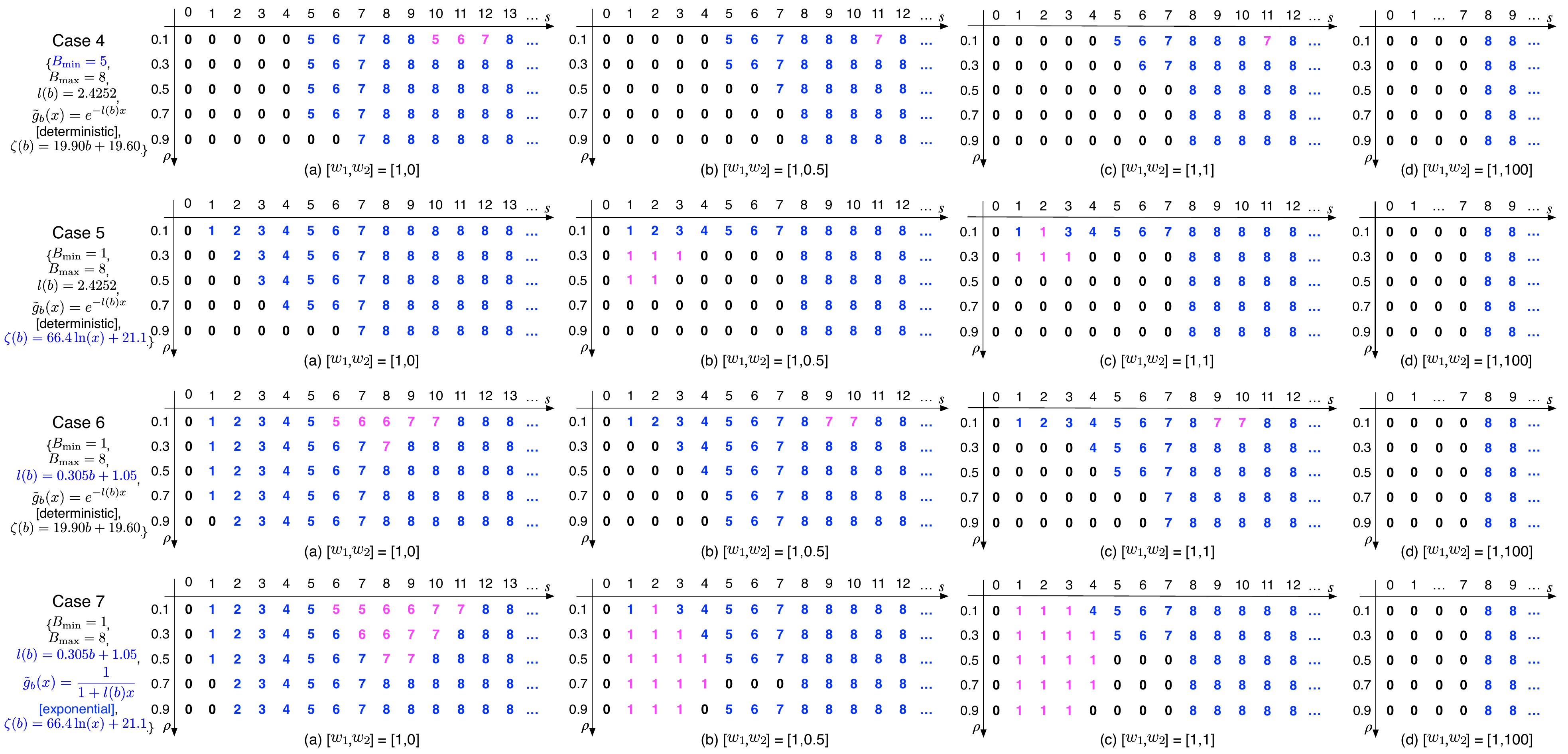}
\caption{The converged SMDP solutions under cases with characteristics such as a minimum batch size greater than 1, a nonlinear energy consumption function, or size-dependent batch service time. The control
limit structure may not be applicable, with the elements that disrupt
the structure highlighted in magenta.} 
\label{policy_view_outcase}
\end{figure*}

We construct four scenarios  named Cases 4-7, based on the basic scenario.
The maximum batch size $B_{\max}$ is set as 8.
The converged SMDP solutions under Cases 4-7 are demonstrated in Fig.~\ref{policy_view_outcase}.
In Cases 4-6, which violate Assumptions 2, 3, and 1, respectively, it is observed that not all SMDP solutions adhere to the control limit structure. 
The elements that disrupt the control limit structure are highlighted in magenta.
In Case 4, where the minimum batch size $B_{\min}$ is set to 5, the SMDP solutions do not merely adjust the control limits $Q$ of Case 1 to $\max(Q,5)$. 
For example, when $\rho=0.1$ and $w_2=0, 0.5$ and $1$, there are states that exceed $B_{\max}$ with corresponding actions smaller than $B_{\max}$, which contradicts the definition of control limit policies.
In Case 5, some solutions feature more than one threshold dividing the actions between ``wait" ($a(s)=0$) and ``serve" ($a(s)>0$).
In Case 6, there are actions involving serving a batch smaller than the maximum available size.
This can be attributed to the possibility of forming a larger batch when the newly initiated service is completed, with some requests remaining in the buffer. 
In the more general scenario, Case 7, there are even more instances where the solutions deviate from the control limit structure.
    
\section{Comparison with Approximate Iteration Algorithms}\label{appE}
\begin{table*}[h!]
\scriptsize
\caption{Evaluation of approximations under different iteration algorithms.}
\centering
\begin{tabular}
{|m{0.9cm}<{\centering}
m{0.85cm}<{\centering}
m{1.5cm}<{\centering}|
m{0.9cm}<{\centering}
m{0.85cm}<{\centering}
m{1.5cm}<{\centering}|
m{0.9cm}<{\centering}
m{0.85cm}<{\centering}
m{0.85cm}<{\centering}|
m{0.9cm}<{\centering}
m{0.85cm}<{\centering}
m{0.85cm}<{\centering}|}
\hline 
\multicolumn{3}{|c|}{\textbf{RVI}} & \multicolumn{3}{c|}{\textbf{RVI}} & \multicolumn{3}{c|}{{\textbf{AVI}}}
& \multicolumn{3}{c|}{{\textbf{API}}}\\
\multicolumn{3}{|c|}{($c_{\mathrm{o}}=0,s_{\max}=160$)} & \multicolumn{3}{c|}{($c_{\mathrm{o}}=100,s_{\max}=160$)} & \multicolumn{3}{c|}{(Scheme I in \cite{thomas1985finite})}
& \multicolumn{3}{c|}{(Scheme IV in \cite{thomas1985finite})}\\
\hline 
CPU time [s] &
$\hat{g}^{{\pi}}$ & 
$\Delta^{\pi}$ &
CPU time [s] &
$\hat{g}^{{\pi}}$ &
$\Delta^{\pi}$ &
CPU time [s] &
$\hat{g}^{{\pi}}$ & $\hat{g}^{{\pi}_{\text{trunc}}}$ &
CPU time [s] &
$\hat{g}^{{\pi}}$ & $\hat{g}^{{\pi}_{\text{trunc}}}$ 
\\
\hline 
$0.73$ & $108.14$ & $108.14$ &
$0.49$ & $114.46$ & $1.51 \times 10^{-14}$ &
$0.73$ & $197.33$ & $197.33$ & $0.79$ & $909.73$ & $208.14$
\\
$1.81$ & $108.14$ & $108.14$ &
$0.61$ & $44.47$ & $8.70 \times 10^{-15}$ &
$9.12$ & $383.20$ & $208.14$ & $3.46$ & $909.73$ & $208.14$
\\
$1.93$ & $38.86$ & $3.63 \times 10^{-15}$ &
$0.73$ & $38.86$ & $1.42 \times 10^{-15}$ &
$9.36$ & $386.58$ & $42.53$ & $3.94$ & $909.73$ & $42.53$
\\
$3.02$ & $38.86$ & $3.63 \times 10^{-15}$ &
$2.06$ & $38.86$ & $1.48 \times 10^{-15}$ &
$12.220$ & $417.00$ & $42.53$ & $5.25$ & $909.73$ & $42.53$
\\
\hline
\end{tabular}
\label{tab:cputime}
\end{table*}

We compare the proposed finite state approximation scheme with two classical approximate iteration algorithms.
Scheme I in \cite{thomas1985finite} (also Scheme II in \cite{white1982finite}) is a typical approximate value iteration algorithm (abbreviated as AVI in this paper).
Scheme IV in \cite{thomas1985finite}  is an approximate policy iteration algorithm (abbreviated as API in this paper) that incorporates the aforementioned AVI algorithm in its inner loop.
AVI and API algorithms are applied to solve the infinite state discrete-time MDP directly associated with the original SMDP.


The experiments are conducted in the basic scenario, with $\rho=0.5$ and $[w_1,w_2]=[1,1]$.
In our proposed schemes, we set $s_{\max}$ to $160$ and $c_{\mathrm{o}}$ to $0$ and $100$, respectively.
For all value iterations, we initialize the value functions to zero.
The API algorithm has an initial policy set as $a(s)=0$ for all $s$.
The number of inner iterations is set to $20*i$ in the $i$th ($i=1,2,...$) outer iteration loop.
The algorithms are implemented in MATLAB\_R2021b and executed on a MacBook Air (M2, 2022).
The Apple M2 chip is equipped with an 8-core CPU comprising four performance cores and four efficiency cores.

In Table~\ref{tab:cputime}, we demonstrate the change in the evaluated average cost $\hat{g}^{\pi}$ with the exact CPU execution time (averaged over 11 runs) under different schemes.
We also record the evolution of $\Delta^{\pi}$ under the proposed scheme (referred to as RVI in Table~\ref{tab:cputime}).
In both AVI and API, the state space consistently expands with each iteration, while the exact number of iterations used for computing the policy on a state $s$ decreases with the increasing value of $s$.
Consequently, the latter part of the computed policy (the policy computed for relatively large states) does not converge very effectively.
Therefore, we introduce $\pi_{\text{trunc}}$ to truncate and only maintain the policy on the state space $\mathcal{S}_{\text{trunc}}=\{0,1,2,\ldots,160,161\}$, whose  cardinality is the same as the state space $\hat{\mathcal{S}}$ of the proposed scheme.
We use $\hat{g}^{\pi_{\text{trunc}}}$ to denote
the average cost of $\pi_{\text{trunc}}$ evaluated on  $\mathcal{S}_{\text{trunc}}$.

Table~\ref{tab:cputime} shows that the proposed methods with $c_{\mathrm{o}}=0$ and $c_{\mathrm{o}}=100$ start converging at approximately $1.93$ seconds and $0.73$ seconds, respectively. 
The converged $\Delta^{\pi}$ values are on the order of $10^{-15}$, ensuring high approximation accuracy. 
In contrast, AVI and API do not achieve convergence for the entire computed policy within the demonstrated CPU times, but their truncated policies converge around $9.36$ seconds and $3.94$ seconds, respectively. 
Both RVI schemes demonstrate faster convergence than AVI and API, with RVI ($c_{\mathrm{o}}=100$) exhibiting a notable advantage over other schemes.
Furthermore, the converged average cost $\hat{g}^{\pi}$ in RVI is $38.86$, which is lower than the converged average cost $\hat{g}^{\pi_{\text{trunc}}}=42.53$ in AVI and API algorithms. 
This suggests that the converged policies obtained through the proposed schemes provide better approximations to optimal policies than those obtained from AVI and API algorithms.
Moreover, approximate iteration algorithms may encompass a very large state space as the number of iterations increases, which raises challenges in complexity and numerical stability.
\bibliographystyle{IEEEtran}  
\bibliography{zotero_bib} 

 





\end{document}